\documentclass[prc,twocolumn,showpacs,superscriptaddress,preprintnumbers,amsmath,amssymb,nofootinbib]{revtex4-2}
\usepackage{bm}
\usepackage{dcolumn}
\usepackage{multirow}
\usepackage{ifpdf}
\usepackage{url}
\usepackage[utf8]{inputenc}

\ifpdf
  \usepackage{graphicx}     
  \usepackage{hyperref}
\else     
  \usepackage[dvipdfmx]{graphicx}     
  \usepackage[dvipdfmx]{hyperref}
\fi

\newcommand{\vstrut}[2][0em]{\vphantom{\rule[#1]{0pt}{#2}}}
\newcommand{\thatis}{i.e.}

\DeclareMathOperator{\diag}{diag}

\newcommand{\etas}{\eta_\mathrm{s}}
\newcommand{\tauR}{\tau_\mathrm{R}}
\newcommand{\tauI}{\tau_\mathrm{in}}
\newcommand{\Tc}{T_\mathrm{c}}
\newcommand{\chiB}{\chi^\mathrm{tot}}

\newcommand{\fm}{~\text{fm}}
\newcommand{\MeV}{~\text{MeV}}

\usepackage{ulem}  
\ifpdf
  \usepackage{color}
\else
  \usepackage[dvipdfmx]{color} 
\fi

\hypersetup{
colorlinks=true,
linkcolor=blue,
citecolor=blue,
urlcolor=blue
}

\begin{document}
\title{Dynamical evolution of critical fluctuations with second-order baryon
diffusion coupled to chiral condensate}

\author{Azumi Sakai}
\email{azumi-sakai@hiroshima-u.ac.jp}
\affiliation{%
Physics Program, Graduate School of Advanced Science and Engineering,
Hiroshima University, Higashi-Hiroshima, Hiroshima 739-8526, Japan
}
\affiliation{%
Department of Physics, Sophia University, Chiyoda, Tokyo 102-8554, Japan}
\author{Koichi Murase}
\email{koichi.murase@yukawa.kyoto-u.ac.jp}
\affiliation{Department of Physics, Tokyo Metropolitan University, Hachioji, Tokyo 192-0397, Japan}
\affiliation{Yukawa Institute for Theoretical Physics, Kyoto University, Sakyo, Kyoto 606-8502, Japan}
\author{Hirotsugu Fujii}
\email{hr-fujii@nishogakusha-u.ac.jp}
\affiliation{%
International Politics and Economics, Nishogakusha University, Chiyoda, Tokyo 102-8336, Japan}
\affiliation{%
Institute of Physics, Unviversity of Tokyo, Meguro, Tokyo 153-8902, Japan}
\author{Tetsufumi Hirano}
\email{hirano@sophia.ac.jp}
\affiliation{%
Department of Physics, Sophia University, Chiyoda, Tokyo 102-8554, Japan}

\date{\today}

\begin{abstract}
We develop a dynamical model to describe critical fluctuations in heavy-ion
collisions, incorporating the baryon diffusion current and chiral condensate
as dynamical degrees of freedom, to address their nontrivial scale separation.
The model couples fluctuations of the chiral condensate $\sigma$ with baryon density
fluctuations $n$ and the diffusion current $\nu$ based on a second-order
diffusion equation with a finite relaxation time of the baryon diffusion $\tauR$.
We analyze
the spacetime evolution and these correlation functions of the fluctuations in
one-dimensionally expanding background.
We confirm that an appropriate relaxation time $\tauR$ ensures causality.
We show that
propagating waves with finite $\tauR$ split into two modes at the
critical temperature due to
a rapid change of kinetic coefficients.
In the correlation functions,
we find that dynamical $\sigma$ blurs the structure and peak
around the critical temperature.
With finite $\tauR$, the
effect of the critical fluctuations persists longer into the later stages of the evolution.
These findings suggest importance of dynamical effects of the chiral condensate
and baryon diffusion current in identifying critical-point signals in heavy-ion
collisions, where the scale separation is nontrivial.
\end{abstract}

\maketitle

\section{Introduction}
\label{sec:intro}
The exploration of the phase structure of quantum chromodynamics (QCD) has
been an attractive topic in fundamental physics.
The first-principles lattice QCD simulations have shown that the transition
from the hadron gas at low temperature to the quark-gluon plasma (QGP) at high
temperature is a smooth crossover at vanishing baryon chemical potential
$\mu_B$~\cite{Aoki:2006we}.  However, the finite $\mu_B$ region is still a
nontrivial target for
lattice simulations based on the Monte-Carlo integration
due to the notorious sign problem.
Studies based on effective models and
lattice QCD suggest that the crossover may turn into a first-order phase
transition at a critical point as we increase $\mu_B$~\cite{Asakawa:1989bq,
Barducci:1989wi, Halasz:1998qr, Berges:1998rc, Scavenius:2000qd, Fodor:2001pe,
Antoniou:2002xq, Hatta:2002sj}, but neither the location nor the existence of
the QCD critical point has been established so far.

To explore the QCD phase structure experimentally,
we create hot and dense QCD matter transiently by
high-energy heavy-ion collisions.
Since the early 2000s, when the formation of a strongly correlated (and strongly coupled) QGP was experimentally
suggested, the properties of this matter have been
extensively studied through both experimental and theoretical efforts (see,
for example, Refs.~\cite{Akiba:2015jwa, Busza:2018rrf, Arslandok:2023utm}  for summaries
of progress and future directions).
Today, one of the major interests in the heavy-ion collisions is
the search for experimental signatures of the QCD critical
point~\cite{Bzdak:2019pkr, Bluhm:2020mpc, Pandav:2022xxx}.
Systems with finite chemical potential
$\mu_B$ can be generated by lowering the beam energy in heavy-ion collisions.
Such experiments are performed at RHIC Beam Energy Scan (BES)
program~\cite{Odyniec:2013aaa}, NA61/SHINE~\cite{NA61:2014lfx}, and
HADES~\cite{HADES:2019auv}.  Also, new experiments are planned at
FAIR~\cite{Friman:2011zz}, NICA~\cite{MPD:2022qhn}, HIAF~\cite{Yang:2013yeb,
Zhou:2022pxl}, and J-PARC~\cite{Sako:2014fha, J-PARCHeavy-Ion:2016ikk}\@.

Fluctuation observables~\cite{Jeon:2003gk, Koch:2008ia, Asakawa:2015ybt,
Luo:2017faz}, which are related to the baryon-number
susceptibilities~\cite{McLerran:1987pz, Gottlieb:1987ac, Gottlieb:1988cq,
Gavai:1989ce}, are key to probing the QCD critical point in the high-energy
heavy-ion collisions.
Since the fluctuations in conserved charges are enhanced near the critical
point, theoretical studies have suggested ways to identify a signal of the
critical fluctuations by measuring the fluctuation observables in heavy-ion
collisions~\cite{Stephanov:1998dy, Stephanov:1999zu, Hatta:2003wn,
Kitazawa:2012at, Sun:2017xrx}.
%
To quantitatively analyze the signals of the QCD critical point in
experimental observables,
  we need a fully dynamical framework capable of describing
the nontrivial dynamics of phase transition (see Refs.~\cite{Bluhm:2020mpc,
Wu:2021xgu} for reviews).
One of the experimental challenges is the fact
that we cannot directly measure the conserved charge fluctuations inside the transient matter,
particularly at the critical temperature.
Here, we need a dynamical framework 
to trace back the dynamics from the final-state spectra of the observed particles,
in order to extract information at the transition temperature.

In this work,
we investigate a 1+1 dimensionally expanding system
in which the baryon number $n$ couples to other modes.
The baryon number diffusion current $\nu$ and the scalar density $\sigma$
are often neglected as fast modes in idealized theoretical setups.
However, heavy-ion collisions are violent events where 
distinguishing the expected critical fluctuations from other fluctuations
will be non-trivial, and the coupling among them may become significant.
Therefore, we focus on how these fast modes influence the evolution of critical baryon number fluctuation in collision events.

Let us briefly review here the current status of dynamical modeling for
  heavy-ion collisions in the context of the critical point search.
Relativistic hydrodynamics, which is the low-energy effective theory of slow
modes of the system, lays the basis of the dynamical models to describe the
spacetime evolution of the QGP created in heavy-ion collisions.  The critical
fluctuations appear as dynamical fluctuations in relativistic hydrodynamics.
Those dynamical fluctuations~\cite{Nahrgang:2011mg, Kapusta:2011gt,
Murase:2013tma} are studied mainly by two approaches: One is stochastic
hydrodynamics~\cite{Nahrgang:2011vn, Murase:2016rhl, Murase:2019cwc,
Bluhm:2018plm, Singh:2018dpk, Chattopadhyay:2023jfm, Attieh:2024vbl} to
directly describe each realization of the fluctuations.  The other is the
Fokker--Planck-type approach, which effectively describes the evolution of the
fluctuation distribution through the $n$-point correlations of the
fluctuations, such as the hydro-kinetic theory and
Hydro+~\cite{Akamatsu:2016llw, Stephanov:2017ghc, Rajagopal:2019xwg,
An:2019osr, An:2019csj, Jiang:2023slb}.

The normal modes of the QCD critical point in the static and uniform background
is in general given by a linear mixing of the chiral condensate $\sigma$, the
baryon charge density $n$, and the energy density $\epsilon$~\cite{Son:2004iv,
Fujii:2004jt}.  Although the quantitative details of the coupling of $\sigma$,
$n$, and $\epsilon$ is not known, we may construct effective models to
investigate the nature of the critical dynamics in the heavy-ion collisions and
constrain the effective models.
The dynamical universality class of QCD critical point
is expected to be model H in the
classification by Hohenberg and Halperin~\cite{Hohenberg:1977ym}, where the
coupling to the momentum density is considered.  This type of model is recently
started to be analyzed~\cite{Chattopadhyay:2024jlh}, but historically,
simplified models have been used to study the qualitative nature of the
dynamics.
One simplified approach to follow the critical dynamics in the heavy-ion
collisions is to study the dynamics of the chiral condensate $\sigma$ or the
chiral fields $(\sigma, \vec\pi)$~\cite{Nahrgang:2011mg,Nahrgang:2011vn,
Mukherjee:2015swa, Wu:2018twy, Tang:2023zvj, Jiang:2023slb, Attieh:2024vbl}.
Instead of considering the finite baryon density $n$, based on the linear sigma
model, the coupling $g$ between the chiral fields and quarks can be varied to
switch between a crossover, critical point, and first-order phase
transition~\cite{Nahrgang:2011mg,Nahrgang:2011vn}.  This motivates analyses
using the dynamical models with only $\sigma$ (model A) or the $\sigma$ field
coupled to hydrodynamics.
Another simplified approach is to follow the dynamics of the baryon-number
density fluctuations $n$.  The long-range dynamics of the system is governed by
slow modes, which is the conserved-charge density $n$ in the present
case~\cite{Hohenberg:1977ym, Son:2004iv, Fujii:2004jt} where the dynamical
degrees of freedom of the fast mode $\sigma$ becomes irrelevant.  To analyze
the dynamics near the critical point in the heavy-ion collisions, it is useful
to limit the dynamics with dynamical $n$.  This type of model is associated
with model B\@.  The dynamics of baryon number density fluctuation $n$ near the
critical point has been studied by stochastic diffusion equations in
one-dimensionally expanding systems~\cite{Sakaida:2017rtj, Nahrgang:2017hkh,
Nahrgang:2018afz, Bluhm:2019yfb, Wu:2019qfz, Pihan:2022xcl}.

Those existing studies have mainly focused on simplified setups in a
one-dimensionally expanding system or the coupling with ideal hydrodynamics.
However, a modern quantitative description of the heavy-ion collision reactions
requires more realistic dynamical models~\cite{Song:2010mg, Hirano:2012kj},
which are numerical frameworks including an initialization model, relativistic
dissipative hydrodynamics with viscosity and diffusion, and the subsequent
hadronic rescatterings.  The community is moving forward to integrate the
models for critical dynamics in realistic dynamical models and analyze the
related experimental observables quantitatively~\cite{Bluhm:2020mpc}.  For
relativistic dissipative hydrodynamics in such numerical frameworks,
second-order M\"uller--Israel--Stewart (MIS)~\cite{Mueller:1967m1,
Israel:1976efz, Israel:1979wp}-type theories have been traditionally used to
avoid acausal and unstable spurious modes in the first-order Navier-Stokes
theory with naive truncation of the derivative expansion~\cite{Hiscock:1983zz,
Hiscock:1985zz}.  It should be noted here that the first-order causal and
stable theories~\cite{Bemfica:2017wps, Kovtun:2019hdm} by Bemfica, Disconzi,
Noronha, and Kovtun (BDNK)
have also been the subject of recent active theoretical discussion,
yet the dynamical models based on them are still under
development~\cite{Noronha:2024dtq}.  The MIS-type theories add new dynamical
degrees for the deviation of the dissipative currents from their first-order
contribution.  These new modes are fast modes following the relaxation-type
equation of motion with finite relaxation times, which is related to the
microscopic scale.  In heavy-ion collisions, the relaxation time $\tauR$ is
typically of the order $\tauR \sim \eta/(e+P) \sim 1/T \sim 1\fm$ (with $\eta$
and $P$ being the viscous coefficient and the pressure), which is not well
separated from the lifetime and the size of the total system, $1$--$10\fm$.
Therefore, these fast modes are not just important for the stability and
causality of the numerical simulations but also physically non-negligible
because of the insufficient scale separation in the heavy-ion collisions.

In existing studies of the critical dynamics based on model B, only $n$ has
been considered for simplification and also for the presumed irrelevance of the fast
mode $\sigma$ in the long-range phenomena~\cite{Son:2004iv, Fujii:2004jt}.
However, as already discussed, the scale separation is not well established in the heavy-ion
collisions.  It remains nontrivial whether we may reduce a
dynamical degree in the equations of motion for the coupled $n$--$\sigma$
system~\cite{Son:2004iv, Fujii:2004jt} in the context of heavy-ion
collisions.
In particular, given that existing second-order relativistic hydrodynamics already solves the fast modes of the
dissipative currents as dynamical degrees of freedom, it is
natural to treat both fast modes, $\sigma$ and the dissipative currents $\nu$,
on an equal footing.

The purpose of this paper is to formulate and analyze the
$n$--$\sigma$--$\nu$ system, namely the second-order diffusion equation system
of the baryon density $n$ and the diffusion current $\nu$, coupled to the
relaxation equation of the chiral condensate $\sigma$.  This study intends to
lay the basis for future implementations and analyses based on realistic
dynamical models of the heavy-ion collisions.

This paper is organized as follows: We first extend the $n$--$\sigma$ system to
the $n$--$\sigma$--$\nu$ system with the analysis of its behavior and derive
the equations of the (1+1)d evolution on top of the one-dimensional Bjorken
expansion in Sec.~\ref{sec:model}\@.  We next introduce the parameter setups
for the $n$--$\sigma$--$\nu$ system in an expanding background that passes near
the critical point in Sec.~\ref{sec:parameters}\@.  We implement the
$n$--$\sigma$--$\nu$ system in a numerical code and compare the Green's
functions of the $n$--$\sigma$ and $n$--$\sigma$--$\nu$ systems to understand
the effect of the relaxation time of the baryon diffusion $\tauR$ in Sec.~\ref{sec:Green}\@.  We
also analyze the effects on the two-point correlations and the time dependence
of the peak at $\Delta\etas = 0$, which is related to the second-order cumulants
measured in experiments in Sec.~\ref{sec:Corr}\@.  Finally,
Sec.~\ref{sec:conclusion} is devoted to the summary and outlook.
We summarize the choice of parameters for susceptibilities and diffusion constant
in Appendix~\ref{app:sakaida}\@.
We also give our
numerical scheme in Appendix~\ref{app:numerical}\@.
We use the
natural units, $\hbar = c = k_\mathrm{B} = 1$, and the sign convention of the metric,
$g_{\mu\nu} = \diag(+, -, -, -)$, throughout this paper.

\section{Model}
\label{sec:model}
In this section, we introduce the $n$--$\sigma$--$\nu$ system with $\nu$ being
the baryon diffusion current.  After reviewing the $n$--$\sigma$ system in
Sec.~\ref{sec:model.n-sigma}, we formulate the $n$--$\sigma$--$\nu$ system and
discuss the dispersion relations of its normal modes
in Sec.~\ref{sec:model.n-sigma-nu}\@.
In Sec.~\ref{sec:model.n-nu}, we consider the $n$--$\nu$ and $n$ systems as
limiting cases of the $n$--$\sigma$--$\nu$ system and summarize the features of
the introduced four systems. Finally, in
Sec.~\ref{sec:model.n-sigma-nu.expanding}, we derive the evolution equations for
the $n$--$\sigma$--$\nu$ system in a 1+1 dimensionally expanding system.

\subsection{\texorpdfstring{$n$--$\sigma$}{n--σ} system}
\label{sec:model.n-sigma}
We consider the fluctuations of the baryon number density $ \delta n := \sum_f
\tfrac{1}{3}\langle \bar q_f \gamma^0 q_f \rangle - n_0$ and the scalar density
$\delta \sigma := \sum_f\langle \bar q_f q_f \rangle -\sigma_0$, where $f$
specifies a quark flavor. The symbol $\langle\cdots\rangle$ denotes the ensemble average
with source fields (i.e., the Lagrange-multiplier fields added in the density operator)
coupled to the operators $\bar q_f \gamma^0 q_f$ and $\bar q_f q_f$,
and $n_0$ and $\sigma_0$ are
the equilibrium values with vanishing source fields.  For simplicity, we
denote these fluctuations by $n$ and $\sigma$ omitting the symbol $\delta$
hereafter.  We do not consider their coupling to the energy and momentum in
this study.

We may write the fluctuation part of the Ginzburg--Landau functional $F$ up to
the quadratic order with respect to the fluctuations~\cite{Son:2004iv}:
\begin{subequations}
\begin{align}
F[n, \sigma] &= \int d^3\bm{x}
\Bigl[\frac{a}{2}(\partial_i \sigma)^2 +b\partial_i \sigma \partial_i n
  \notag \\ &\quad
  + \frac{c}{2}(\partial_i n)^2
  +V(n, \sigma)\Bigr]\;, \label{eq:GL_F}\\
V(n, \sigma) &= \frac{A}{2} \sigma^2  + B\sigma  n + \frac{C}{2} n^2\;, \label{eq:GL_V}
\end{align}
\end{subequations}
with the expansion coefficients\footnote{
  The definition of those expansion coefficients may depend on the literature.
  For example, Ref.~\cite{Fujii:2004jt} gives the coefficients by expanding
  $\Omega = \beta F$ instead of $F$, which results in the difference in the
  coefficient definitions by the factor $\beta$ compared to ours.
}
being $a$, $b$, $c$, $A$, $B$, and $C$.  We kept only the potential part, $V(\sigma, n)$,
to work at the leading order in the spatial derivatives in this work.  The
susceptibilities of $n$ and $\sigma$ in the equilibrium distribution $\propto
\exp(-\beta F)$ are
\begin{align}
  \chi_{nn} = \frac{TA}{\Delta}\;, \quad
  \chi_{n \sigma} = -\frac{TB}{\Delta}\;, \quad
  \chi_{\sigma\sigma} = \frac{TC}{\Delta}\;,
  \label{eq:n-sigma.susceptibilities}
\end{align}
with $\Delta = AC-B^2$.

The stochastic hydrodynamic equations are written as~\cite{Son:2004iv}
\begin{align}
  \frac{\partial}{\partial t}
  \begin{pmatrix}
    n \\ \sigma
  \end{pmatrix} &=
  -\begin{pmatrix}
    -\lambda \nabla^2 & -\tilde \lambda \nabla^2 \\
    -\tilde \lambda \nabla^2 & \Gamma
  \end{pmatrix}
  \begin{pmatrix}
    \frac{\delta F}{\delta n} \\
    \frac{\delta F}{\delta \sigma} \vstrut{1.2em}
  \end{pmatrix}
  + \begin{pmatrix} \xi_n  \\ \xi_\sigma \end{pmatrix}\;,
  \label{eq:critical_fluctuations}
\end{align}
where $\Gamma$, $\lambda$, and $\tilde\lambda$ are the kinetic coefficients,
and the terms $\xi_n$ and $\xi_\sigma$ are Gaussian white noises.

The coefficient matrix in Eq.~\eqref{eq:critical_fluctuations}, also known as
the Onsager matrix, should be symmetric to satisfy Onsagar's reciprocal
relation.  The Onsager matrix should also be positive-semidefinite to respect
the second law of thermodynamics, \thatis, the eigenvalues are required to be
non-negative for thermodynamic stability.  The Onsager matrix is diagonalized
in the Fourier space, and it is positive-semidefinite when the trace
$\lambda\bm{k}^2 + \Gamma$ and the determinant $\lambda\bm{k}^2\Gamma -
\tilde\lambda^2\bm{k}^4$ are both non-negative, which leads to the following
conditions:
\begin{align}
  \lambda,\Gamma &\ge 0\;, & \frac{|\tilde\lambda|}{\sqrt{\Gamma\lambda}} |\bm{k}| \le 1\;.
  \label{eq:onsager-cutoff}
\end{align}
This means that when $\tilde\lambda\ne 0$, a perturbation with a shorter scale
than $|\tilde\lambda|/\sqrt{\Gamma\lambda}$ is unstable.  In other words, the
dynamical equation~\eqref{eq:critical_fluctuations} has an intrinsic cutoff
coming from its truncated structure (see Sec.~IV~D of
Ref.~\cite{Murase:2019cwc} for the general discussion).

The power spectra of the noise terms are given by the fluctuation-dissipation
relations (FDRs):
\begin{subequations}
\begin{align}
\langle \xi_{n}   (x) \xi_{n}   (x') \rangle &= 2T(-\lambda\nabla^2)\delta^4 (x-x') \; ,\label{FDR1}\\
\langle \xi_{n}   (x) \xi_\sigma(x') \rangle &= 2T(-\tilde \lambda\nabla^2)\delta^4 (x-x')\; , \label{FDR2}\\
\langle \xi_\sigma(x) \xi_\sigma(x') \rangle &= 2T\Gamma \delta^4 (x-x')\;. \label{FDR3}
\end{align}
\end{subequations}
FDRs are given by the temperature and the coefficients in the Onsager matrix in
Eq.~\eqref{eq:critical_fluctuations}.  The spatial derivatives $-\nabla^2$ come
from the Onsager matrix, which is also required by the baryon number
conservation: the noise term $\xi_{n}$ needs to be written in the form $\xi_n =
-\nabla\cdot\bm{\xi}_\nu$ so that the dynamical evolution of $n$ keeps the form
of the continuity equation, $\partial n/\partial t = -\nabla\cdot\bm{J}$.

Besides, the autocorrelation of the noise needs to be specified by a
positive-semidefinite function\footnote{
  The matrix $M(x,x')$ as a function of $x$ and $x'$ is said to be
  positive-semidefinite when the following inequality is satisfied for
  any vector function $\bm{v}(x)$:
  \begin{align*}
    \int dx dx'\bm{v}^\dag(x) M(x, x') \bm{v}(x') \ge 0.
  \end{align*}
  The autocorrelation matrix $\langle\bm{\xi}(x)\bm{\xi}^\dag(x')\rangle$ is
  positive-semidefinite because
  \begin{align*}
    \int dx dx'\bm{v}^\dag(x)\langle \bm{\xi}(x)\bm{\xi}^\dag(x')\rangle\bm{v}(x')
    = \Bigl\langle\Bigl|\int dx \bm{v}(x)^\dag\bm{\xi}(x)\Bigr|^2\Bigr\rangle.
  \end{align*}
}. This is ensured by the positive-semidefiniteness of the Onsager matrix.
As higher momentum modes $|\bm{k}| > |\tilde\lambda|/\sqrt{\Gamma\lambda}$ violate
  the condition \eqref{eq:onsager-cutoff} when $\tilde\lambda \ne 0$, we must explicitly introduce
  a momentum cutoff of the delta functions to remain consistent with the physical requirement
  present before the scale separation approximation.

The explicit form of the equations reads
\begin{align}
  \frac{\partial}{\partial t}
  \begin{pmatrix}
    n \\ \sigma
  \end{pmatrix} &=
  -\begin{pmatrix}
    -\lambda \nabla^2 & -\tilde \lambda \nabla^2 \\
    0 & \Gamma
  \end{pmatrix}
  \begin{pmatrix}
    C & B \\ B & A
  \end{pmatrix}
  \begin{pmatrix} n  \\ \sigma \end{pmatrix}
  + \begin{pmatrix} \xi_n  \\ \xi_\sigma \end{pmatrix}\;,
  \label{eq:n-sigma.eom}
\end{align}
up to the leading derivative terms.  We dropped the bottom left component of
the Onsager matrix because it is subdominant compared to $\Gamma$ in the
derivative expansion.

The dispersion relations for Eqs.~\eqref{eq:n-sigma.eom} are found to be
\begin{align}
  \omega_\pm &= -i\frac{u \pm \sqrt{u^2 - 4\Delta\lambda \bm{k}^2 \Gamma}}2\;,
  \label{eq:dispersion.2x2}
\end{align}
where $u = \lambda \bm{k}^2 C + \tilde \lambda \bm{k}^2 B + \Gamma A$.  In the
limit $|\bm{k}| \ll 1$, the modes $\omega_+$ and $\omega_-$ become a diffusive
mode $\omega_D$ and a relaxation mode $\omega_\sigma$, respectively, at the
lowest order of $\bm{k}^2$~\cite{Son:2004iv, Fujii:2004jt, Fujii:2004za}:
\begin{align}
  \omega_D = -i D \bm{k}^2\;, \quad
  \omega_\sigma = -\frac{i}{\tau_\sigma}\;,
\end{align}
where $D := \lambda\Delta/A$ is the diffusion constant, and $\tau_\sigma
:=1/\Gamma A$ is the relaxation time of $\sigma$.  It should be noted that the
aforementioned unstable modes violating Eq.~\eqref{eq:onsager-cutoff} do not
appear due to the absence of the left-bottom component of the Onsager matrix in
Eq.~\eqref{eq:n-sigma.eom}.  The first mode $\omega_D$ is a linear combination
of $n$ and $\sigma$ fluctuations while the other $\omega_\sigma$ is a pure
$\sigma$ fluctuation.  At the critical point where $\Delta=0$, all the
susceptibilities diverge, and the diffusive mode shows the critical slowing
down.

\subsection{\texorpdfstring{$n$--$\sigma$--$\nu$}{n--σ--ν} system with baryon relaxation time}
\label{sec:model.n-sigma-nu}

The $n$--$\sigma$ system~\eqref{eq:n-sigma.eom} is essentially a diffusion
equation for the $n$ part, in which the signal propagation speed is unbounded
and thus acausal.
As we have discussed in
the Introduction, this is unfavorable in realistic
dynamical modeling of the heavy-ion collisions, where we typically employ the
second-order causal hydrodynamics with the finite relaxation times for
dissipative currents. Meanwhile, it remains nontrivial whether the scale
separation in the heavy-ion collisions is sufficiently large so that the
dynamical degrees of fast modes can be eliminated.  In addition to the baryon
density fluctuation $n(x)$, we consider both the chiral condensate fluctuation
$\sigma(x)$ and the baryon diffusion current $\nu^\mu(x)$ as dynamical degrees.

We extend the $n$--$\sigma$ system by adding a dynamical degree for the
second-order correction of the baryon diffusion current $\bm{\nu} = (\nu^1,
\nu^2, \nu^3)^\mathrm{T}$.  It should be noted that we assume a uniform rest background
$u^\mu \equiv (1, 0, 0 ,0)^\mathrm{T}$ so that the dissipative currents only
have spatial components $\nu^i$ ($i=1$, $2$, $3$).  The full equation set is
given by
\begin{subequations}
  \label{eq:n-sigma-nu-eom}
  \begin{align}
    \frac{\partial n}{\partial t}
      &= -\nabla\cdot\bm{\nu}\;,
      \label{eq:n-sigma-nu-eom.1} \\
    \frac{\partial \sigma}{\partial t}
      &= -\Gamma B n -\Gamma A \sigma + \xi_\sigma\;,
      \label{eq:n-sigma-nu-eom.2} \\
    \tauR \frac{\partial \bm{\nu}}{\partial t} + \bm{\nu}
      &= -\nabla \bigl[
        (\tilde \lambda B + \lambda C) n
        + (\tilde \lambda A + \lambda B) \sigma
      \bigr] + \bm{\xi}_\nu\;,
      \label{eq:n-sigma-nu-eom.3}
  \end{align}
\end{subequations}
where the relaxation time of the baryon diffusion current $\tauR$
(which we call \textit{the baryon relaxation time} hereafter)
is introduced as a new transport
coefficient.
The noise term $\bm{\xi}_\nu = (\xi_\nu^1, \xi_\nu^2,
\xi_\nu^3)^\mathrm{T}$ satisfies $\xi_n = -\nabla\cdot\bm{\xi}_\nu$ as
discussed in the previous subsection.  Instead of directly including $\xi_n$ in
the dynamical equation of the baryon density~\eqref{eq:n-sigma-nu-eom.1}, it is
useful to factorize $-\nabla$ and include the noise current $\bm{\xi}_\nu$ in
the dynamical equation~\eqref{eq:n-sigma-nu-eom.3}.

To reproduce the FDRs of the $n$--$\sigma$ system~\eqref{FDR1}--\eqref{FDR3} in
the limit $\tauR\to0$, the FDRs of the $n$--$\sigma$--$\nu$ system are
identified as
\begin{subequations}
\begin{align}
  \langle \xi_\nu^i  (x) \xi_\nu^j  (x') \rangle &= -\Delta^{ij} 2T\lambda\delta^4 (x-x') \;,\label{eq:n-sigma-nu-FDR.1}\\
  \langle \xi_\nu^i  (x) \xi_\sigma (x') \rangle &= -\Delta^{ij} 2T\tilde\lambda \nabla_j\delta^4 (x-x')\;,\\
  \langle \xi_\sigma (x) \xi_\sigma (x') \rangle &= 2T\Gamma \delta^4 (x-x')\;,
\end{align}
\end{subequations}
where the spatial projector $-\Delta^{ij} = u^i u^j - g^{ij} = \diag(1,1,1)$ is
explicitly written for the later use in the transformation to the Milne coordinates in the next
subsection.

The dispersion relations of the normal modes of the
system~\eqref{eq:n-sigma-nu-eom.1}--\eqref{eq:n-sigma-nu-eom.3} (without the
noise terms) are found to be
\begin{align}
  \omega_\sigma = -\frac{i}{\tau_\sigma}\;, \quad\omega_\mathrm{R} = -\frac{i}{\tauR}\;, \quad
  \omega_D  = -i D \bm{k}^2\;,
\end{align}
at a small wave number $|\bm{k}|$.

At a large wave number $|\bm{k}|$, $n$ and $\nu$ mix to form two propagating modes:
\begin{subequations}
\begin{align}
  \omega^2 = \frac{\tilde \lambda B + \lambda C}{\tauR}\bm{k}^2 =:
  v^2 \bm{k}^2\;,
  \label{eq:dispersion.3x3-highk-prop}
\end{align}
where $v$ gives the wave velocity.  The introduction of $\tauR$ leaves the long
wavelength mode unaffected, while making the signal speed of short-wavelength modes finite,
and thus causality is respected.
Note that
the baryon relaxation time $\tauR$ in the
denominator should diverge near the critical point
to keep the wave velocity finite since the kinetic coefficient $\lambda$ in the numerator diverges at the critical point.
We may extend the constant $\tauR$ used in our present study to a temperature-dependent $\tauR$ in a future study,
as associated with the critical slowing down.
The other mode is also modified at large $|\bm{k}|$ as
\begin{align}
  \omega &= -i\frac{\lambda\Gamma\Delta}{\tilde\lambda B + \lambda C}\;.
  \label{eq:dispersion.3x3-highk-prop-2}
\end{align}
\end{subequations}

\subsection{\texorpdfstring{$n$--$\nu$}{n--ν} and \texorpdfstring{$n$}{n} systems
with fast \texorpdfstring{$\sigma$}{σ} integrated}
\label{sec:model.n-nu}

If the timescale of the $\sigma$ dynamics is faster than the typical timescale
governing $n$ and $\nu$, the chiral condensate $\sigma$ can be approximated by its equilibrium value
$\sigma_\mathrm{eq}$ under a given $n$ when considering the dynamics of $n$ and $\nu$.
The equilibrium value of $\sigma$ is
found to be $\sigma = -(B/A)n$ by solving Eq.~\eqref{eq:n-sigma-nu-eom.2} under
the constant $n$ and without noise.  In this limit, the dynamical equations for
the $n$--$\nu$ system are written as
\begin{subequations}
  \label{eq:model.n-nu.eom}
  \begin{align}
    \frac{\partial n}{\partial t} &= -\nabla\cdot\bm{\nu}\;, \label{eq:nB_eta_tau}\\
    \tauR \frac{\partial \bm{\nu}}{\partial t} + \bm{\nu} &=
      -D \nabla n + \bm{\xi}_\nu\;.
      \label{eq:nu_eta_tau}
  \end{align}
\end{subequations}
This system is essentially equivalent to the frameworks of fluctuating
hydrodynamics discussed in Refs.~\cite{Murase:2013tma, Kapusta:2014dja,
Murase:2019cwc}.

The normal mode dispersions are obtained as
\begin{align}
  \omega = \frac{1}{2\tauR} \bigl( -i \pm \sqrt{-1 + 4D\tauR \bm{k}^2} \bigr)\;,
  \label{eq:dispersion.n-nu.general}
\end{align}
which gives rise to $\omega_D = -i D \bm{k}^2$ and $\omega_\mathrm{R}=-i/\tauR$
at small $|\bm{k}|$ and $\omega = \pm \sqrt{D/\tauR}|\bm{k}|$ at large
$|\bm{k}|$.  The critical slowing down shows up in the diffusion mode as the
diffusion constant $D\propto \Delta$ vanishes at the critical point.

When the baryon relaxation time $\tauR$ is negligible,
Eqs.~\eqref{eq:nB_eta_tau} and~\eqref{eq:nu_eta_tau} are further reduced to a
simple stochastic diffusion equation for the $n$ system~\cite{Kapusta:2011gt,
Sakaida:2017rtj}:
\begin{align}
  \frac{\partial n}{\partial t} &= D \nabla^2 n + \xi_n\;.
      \label{eq:model.n.eom}
\end{align}

\begin{table}[htbp]
\caption{Dynamical systems for critical fluctuations}
\label{table}
\centering
\begin{tabular}{c@{\hspace{2em}} c@{\hspace{2em}} c} \hline \hline
System & $\sigma$ relaxation & $\nu$ relaxation \\
\hline
$n$--$\sigma$--$\nu$ & Finite $\tau_\sigma$ & Finite $\tauR$ \\
$n$--$\nu$~\cite{Murase:2013tma, Kapusta:2014dja} & $\tau_\sigma\to 0$ & Finite $\tauR$ \\
$n$--$\sigma$~\cite{Fujii:2004jt, Son:2004iv} & Finite $\tau_\sigma$ & $\tauR \to 0$\\
$n$~\cite{Sakaida:2017rtj, Kapusta:2011gt} & $\tau_\sigma \to 0$ & $\tauR \to 0$\\
\hline \hline
\end{tabular}
\end{table}

The four systems described so far are summarized in Table~\ref{table}\@.
The $n$--$\sigma$--$\nu$ system is the main model in this work,
where we consider the coupling between $n$ and $\sigma$ as well as the finite
baryon relaxation time
$\tauR$.  The other systems may be obtained by considering the vanishing baryon
relaxation time, $\tauR = 0$, and $\sigma$-relaxation time, $\tau_\sigma =
1/\Gamma A = 0$.
The $n$--$\nu$ system essentially discussed in Refs.~\cite{Murase:2013tma,
Kapusta:2014dja} can be obtained by integrating $\sigma$ as a fast mode in the
limit $\tau_\sigma \to 0$.
The $n$--$\sigma$ system discussed in Refs.~\cite{Fujii:2004jt, Son:2004iv} is
obtained by neglecting the finite baryon relaxation time $\tauR$.
The $n$ system discussed in Refs.~\cite{Sakaida:2017rtj, Kapusta:2011gt} is the
case focusing only on the evolution of baryon density $n$ by eliminating
relaxation times of both $\sigma$ and $\bm{\nu}$.

  For critical phenomena in a large system near equilibrium,
  we typically focus on the long-range fluctuations, where $\nabla \sim \bm{k} \ll \tau_\sigma^{-1}, \tau_R^{-1}$.
 In such cases with scale separation,  
  the $n$--$\sigma$--$\nu$ system, Eqs.~\eqref{eq:n-sigma-nu-eom.1}--\eqref{eq:n-sigma-nu-eom.3},
  reduces to the $n$ system, Eq.~\eqref{eq:model.n.eom}.
  However, heavy-ion collision reactions evolve more dynamically with the space- and time-scales
  typically of the order of 1 fm.
  In the next section, we numerically solve the set of evolution equations,
  Eqs.~\eqref{eq:n-sigma-nu-eom.1}--\eqref{eq:n-sigma-nu-eom.3}, under a simple setup
  with the parameter values chosen to reflect the conditions of collision reactions.
  We then compare the numerical results for system evolution with those of three other models:
  the $n$--$\nu$, $n$--$\sigma$, and $n$ systems.
\subsection{(1+1)d \texorpdfstring{$n$--$\sigma$--$\nu$}{n--σ--ν} system in expanding background}
\label{sec:model.n-sigma-nu.expanding}

In the subsequent numerical analyses, we consider (1+1)d spacetime evolution of
fluctuations on top of the 0+1d Bjorken background (\thatis, a longitudinally
expanding solution with boost invariance).  We adopt the $\tau$--$\etas$
coordinates, also known as the (1+1)d Milne coordinates, to specify the time $t
= \tau\cosh\etas$ and the longitudinal position $z = \tau\sinh\etas$.  We
ignore the transverse coordinates assuming that the matter is uniform and has
infinite extension in the transverse direction, \thatis, $\partial_x =
\partial_y = 0$ and $\nu^x = \nu^y = \xi_\nu^x = \xi_\nu^y = 0$.  The
$n$--$\sigma$--$\nu$ dynamical equation
system~\eqref{eq:n-sigma-nu-eom.1}--\eqref{eq:n-sigma-nu-eom.3} is transformed
into the one in the $\tau$--$\etas$ coordinates $(\tau, x, y, \etas)$ using the
metric $g_{\alpha\beta} = \diag(1,-1,-1,-\tau^2)$, and the Christoffel symbols
$\Gamma^\tau{}_{\etas\etas} = \tau$, $\Gamma^{\etas}{}_{\tau\etas} =
\Gamma^{\etas}{}_{\etas\tau} = 1/\tau$, and otherwise,
$\Gamma^\alpha{}_{\beta\gamma} = 0$, with $\alpha, \beta, \gamma = \tau, x, y,
\etas$.  In the general coordinate system, the baryon number
conservation~\eqref{eq:n-sigma-nu-eom.1} can be concisely expressed in terms of
the \textit{vector density} $(\tilde n, \nu)^\mathrm{T} := \sqrt{-g} (n,
\nu^{\etas})^\mathrm{T} = (\tau n, \tau \nu^{\etas})^\mathrm{T}$ with the
Jacobian $\sqrt{-g}=\sqrt{-\det g_{\alpha\beta}}$.  The other
equations~\eqref{eq:n-sigma-nu-eom.2} and~\eqref{eq:n-sigma-nu-eom.3} can be
transformed by the correspondence $\partial/\partial t = u^\mu \partial_\mu$
and $\nabla = -\Delta^{\mu\nu}\partial_\nu$ in the rest background
$u^\alpha=(1,0,0,0)^\mathrm{T}$.  We finally obtain the following set of dynamical
equations in the expanding system:
\begin{subequations}
\begin{align}
  \frac{\partial  \tilde n}{\partial \tau}
    &= -\frac{\partial \nu}{\partial \etas}\;,
    \label{eq:critical_fluctuations_eta_tau1}\\
  \frac{\partial \sigma}{\partial \tau}
    &= -\Gamma A  \sigma - \Gamma B \frac{ \tilde{n}}{\tau} + \xi_\sigma\;,
    \label{eq:critical_fluctuations_eta_tau2}\\
  \tauR \frac{\partial \nu}{\partial \tau} + \nu
    &= -\frac{1}{\tau} \frac{\partial}{\partial\etas}
      \Bigl[
        (\tilde \lambda A + \lambda B) \sigma 
        +(\tilde \lambda B + \lambda C ) \frac{\tilde n}{\tau}
      \Bigr] + \xi_\nu\;,
    \label{eq:critical_fluctuations_eta_tau3}
\end{align}
\end{subequations}
where the noise term for the current density $\nu$ is defined as $\xi_\nu :=
\tau \xi_\nu^{\etas}$.

For simplicity, we have ignored a term $-d\sigma_0/d\tau$ on the right-hand side of
Eq.~\eqref{eq:critical_fluctuations_eta_tau2} coming from the background
evolution of $\sigma_0$.  The term appears because our choice of the dynamical
variables is the fluctuations from the background equilibrium values $\tilde
n_0 = \tau n_0(\tau)$ and $\sigma_0 = \sigma_0(T(\tau), n_0(\tau))$, but the
underlying time evolution is given for the total values $\tilde n_0 + \tilde n$
and $\sigma_0 + \sigma$.  Since the background values can also change as
functions of the time $\tau$, the effect enters the fluctuation part.
Eq.~\eqref{eq:critical_fluctuations_eta_tau1} does not receive the correction
because of the charge conservation $d\tilde n/d\tau = 0$ under the Bjorken
expansion with vanishing background diffusion $\nu_0 = 0$.

The FDRs are expressed in the $\tau$--$\etas$ coordinates as
\begin{subequations}
\begin{align}
  \langle \xi_\sigma(\tau, \etas) \xi_\sigma(\tau', \etas') \rangle
    &= 2T\Gamma \delta(\tau-\tau') \delta_w(\etas-\etas')\;,
    \label{xi_eta_tau1}\\
  \langle \xi_\nu(\tau, \etas) \xi_\nu(\tau', \etas') \rangle &= 2T\lambda
    \delta(\tau-\tau') \delta_w(\etas-\etas')\;,
    \label{xi_eta_tau2}\\
  \langle \xi_\sigma(\tau, \etas) \xi_\nu(\tau', \etas') \rangle &= 2T\tilde\lambda
	\frac{1}{\tau}
	\frac{\partial}{\partial\etas}
    \delta(\tau-\tau')\delta_w(\etas-\etas')\;,
    \label{xi_eta_tau3}
\end{align}
\end{subequations}
where
\begin{align}
  \delta_w(\etas-\etas'; \tau) &:= \frac{1}{\tau S_{\perp}} \frac1{\sqrt{4\pi w^2}}e^{-(\etas-\etas')^2/4w^2},
  \label{eq:mode.n-sigma-nu-expanding.smeared-delta}
\end{align}
is the smeared delta function in the spatial part of the $\tau$--$\etas$
coordinates.  The momentum cutoff of $\xi_\sigma$ and $\xi_\nu$ in the
$\etas$-direction is introduced by the Gaussian smearing with the width $w$,
where we choose $w=0.02$.  This results in the Gaussian shape of the
autocorrelations with the width $\sqrt2 w$.  The transverse area of the system
$S_{\perp}$ is assumed to be $1\fm^2$ for simplicity.  The temperature $T$ and
other coefficients as functions of $T$ are determined by the background
evolution, i.e., we do not consider the temperature modifications by the
fluctuations.

In the following analyses, we also solve the $n$--$\nu$ system in the expanding
background separately, where the dynamical equations are obtained by
substituting its equilibrium value $\sigma = \sigma_\mathrm{eq} := -
(B/A)(\tilde n/\tau)$:
\begin{subequations}
  \label{eq:model.n-nu-expanding.eom}
  \begin{align}
    \frac{\partial  \tilde n}{\partial \tau}
      &= -\frac{\partial \nu}{\partial \etas}\;,
      \label{eq:model.n-nu-expanding.eom1}\\
    \tauR \frac{\partial \nu}{\partial \tau} + \nu
      &= -\frac{1}{\tau} \frac{\partial}{\partial\etas}
        D\frac{ \tilde{n}}{\tau} + \xi_\nu\;.
      \label{eq:model.n-nu-expanding.eom2}
  \end{align}
\end{subequations}

\section{Parameter setup}
\label{sec:parameters}

In this section, we model a set of temperature-dependent parameters for the
dynamical systems introduced in Sec.~\ref{sec:model} for numerical analyses.
We assume a specific trajectory in $(T, \mu)$ plane (See Appendix~\ref{app:sakaida}),
where $\mu$ is the baryon chemical potential.
We introduce the temperature
dependence of the coefficients $(A, B, C, \Gamma, \lambda, \tilde\lambda)$
by using the baryon-number susceptibility $\chi(T)$
and the baryon diffusion constant
$D(T)$ as constraints.

First, we specify a background evolution of the Bjorken
expansion~\cite{Bjorken:1982qr} following Ref.~\cite{Sakaida:2017rtj}.  We
parametrize the time dependence of the background temperature in the form,
\begin{align}
  T(\tau) &= T_0 \left( \frac{\tau_0}{\tau} \right)^{c_\mathrm{s}^2}\;.
  \label{eq:param.temperature}
\end{align}
This is motivated by the solution to ideal hydrodynamics in the Bjorken
expansion with the equation of state $p = c_\mathrm{s}^2 e$.  We follow
Ref.~\cite{Sakaida:2017rtj} for the initial time $\tau_0 = 1.0\fm$, the initial
temperature $T_0 = 220\MeV$, and the sound velocity $c_\mathrm{s}^2 = 0.15$.

We have seven free parameters in the $n$--$\sigma$--$\nu$
system, Eqs.~\eqref{eq:critical_fluctuations_eta_tau1}--\eqref{eq:critical_fluctuations_eta_tau3}:
the potential parameters $(A, B, C)$, the kinetic coefficients $(\Gamma,
\lambda, \tilde\lambda)$, and the baryon relaxation time $\tauR$.  In general,
those parameters are given as functions of $T$ and $\mu$, thus can change
during the evolution through the temperature
evolution by Eq.~\eqref{eq:param.temperature}.
In the present setup, we consider the parameter values of the order of
$1~\text{fm}$ to be consistent with Ref.~\cite{Sakaida:2017rtj}.  In
particular, the relaxation times of two fast modes, $\tauR$ and $\tau_\sigma$,
are intentionally chosen to be the same order since we do not have a priori
knowledge that either one is larger than the other.

The parameters $A$, $B$, and $C$ are determined by the relation to the
equilibrium susceptibility $\chi$.  The full baryon number susceptibility
$\chiB$ (including the critical divergence at the critical temperature $\Tc$)
and its regular part $\chi^\mathrm{reg}$ satisfy the following relations:
\begin{align}
  \frac{TA}{\Delta} &= \chiB = \chi^\mathrm{cr}(T) + \chi^\mathrm{reg}(T)\;, \label{eq:param.A-in-chiB} \\
  \frac{T}{C} &= \chi^\mathrm{reg}(T)\;.
  \label{eq.param.C}
\end{align}
The critical divergence of $\chiB$ at the critical temperature $\Tc$ implies
the vanishing denominator $\Delta = AC - B^2 = 0$, which leads to a flat
direction in the potential landscape $V(n,\sigma)$.  We need one additional
constraint to determine the coefficients $(A,B,C)$, where we simply assume $B =
1\fm^2$ in the present study.

The kinetic coefficient, $\lambda$, is determined by its relation to the
diffusion constant,
\begin{align}
 \frac{\lambda \Delta}{A} = D(T)\;.
  \label{eq:param.D}
\end{align}
To avoid the broken positive-semidefiniteness of the FDR for large $\bm{k}^2$
modes, and also for the technical simplicity in generating the noises with the
derivative in the FDR~\eqref{xi_eta_tau2}, we assume vanishing
$\tilde\lambda=0$.
For the remaining kinetic coefficient, $\Gamma$, we have no stringent
condition.  For the first trial, we fix the $\sigma$-relaxation time for the
$n$--$\sigma$--$\nu$ system as $\tau_\sigma = 1/\Gamma A = 1\fm$.

In this study, we follow the existing study~\cite{Sakaida:2017rtj} for the
temperature dependences of the susceptibilities $\chiB(T)$ and
$\chi^\mathrm{reg}(T)$, and the diffusion constant $D(T)$, with the choices of
the critical temperature $\Tc=160\MeV$, the singular-part coefficient $c_\mathrm{c}=4$,
and the evolution trajectory $r=10^{-4}$.  In Ref.~\cite{Sakaida:2017rtj}, the
singular part of the critical susceptibility is determined by a mapping of
$(\mu, T)$ to $(r, H)$ in the Ising model, where $r$ and $H$ are the reduced temperature and the magnetic field,
respectively, at the Ising side.  The trajectory $r=10^{-4}$ in the $(r, H)$
plane was chosen so that the system crosses the critical point with a small
offset to avoid the singularity in the numerical calculation.  The singular
part of the diffusion constant is determined based on the universality class of
the dynamic critical behavior of the liquid-gas transition~\cite{Son:2004iv,
Fujii:2004jt, Fujii:2004za}.  The regular parts of the susceptibility and the
diffusion constants are constructed by smoothly connecting the hadron and QGP
values.  Explicit expressions of $\chiB(T)$, $\chi^\mathrm{reg}(T)$, and $D(T)$
given in Ref.~\cite{Sakaida:2017rtj} are recapitulated in
Appendix~\ref{app:sakaida} with additional details.

\begin{figure}[htbp]
  \centering
  \includegraphics[width=0.45\textwidth]{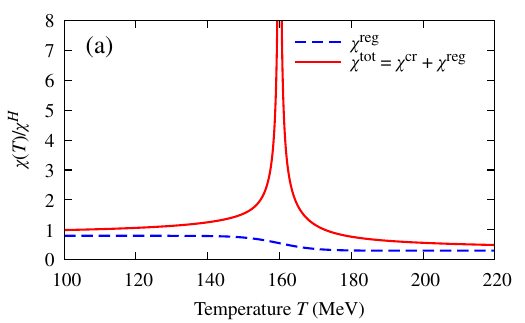}\\
  \includegraphics[width=0.45\textwidth]{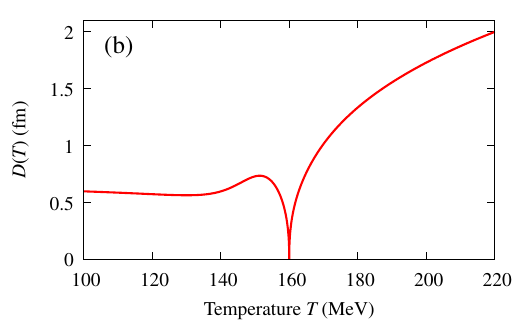}
  \caption{(a) Temperature dependence of susceptibility $\chiB(T)$ and
  $\chi^\mathrm{reg}(T)$.
The blue dashed line and the solid red line show $\chi^\mathrm{reg}(T)$ and $\chiB(T)$,
respectively.
(b) Temperature dependence of diffusion constant $D(T)$ with $c_\mathrm{c} = 4$ and
  $r = 10^{-4}$.}
  \label{fig:kai_and_D}
\end{figure}

Figure~\ref{fig:kai_and_D} shows the temperature dependence of $\chiB(T)$,
$\chi^\mathrm{reg}(T)$, and $D(T)$. These behaviors are qualitatively similar to
those in Fig.~2 of Ref.~\cite{Sakaida:2017rtj}, but there are two differences
to be noted.  First, $\chi^\mathrm{reg}$ in Fig.~\ref{fig:kai_and_D} (a) is
different from ``reg'' in Fig.~2 of Ref.~\cite{Sakaida:2017rtj} because the
former shows the regular part in the total susceptibility, while the latter
shows the total susceptibility when there was no critical contribution.  More
specifically, $\chi^\mathrm{reg}$ in Fig.~\ref{fig:kai_and_D} (a) uses the
parameters determined by Eqs.~\eqref{eq:def.chiH0} and~\eqref{eq:def.chiQ0},
while $\chi^\mathrm{reg}$ in Fig.~2 of Ref.~\cite{Sakaida:2017rtj} uses the
parameters $\chi^\mathrm{H}_0 = \chi^\mathrm{H}$ and $\chi^\mathrm{Q}_0 =
\chi^\mathrm{Q}$ assuming $\chi^\mathrm{cr} = 0$.  The diffusion constant
$D(T)$ in Fig.~\ref{fig:kai_and_D} (b) differs from Fig.~2 of
Ref.~\cite{Sakaida:2017rtj} because we use $c_\mathrm{c}=4$ in Fig.~\ref{fig:kai_and_D} (b)
while a different parameter
$c_\mathrm{c} = 1/2$ is used in Fig.~2 of Ref.~\cite{Sakaida:2017rtj}.

\begin{figure}[htbp]
  \centering
  \includegraphics[width=0.45\textwidth]{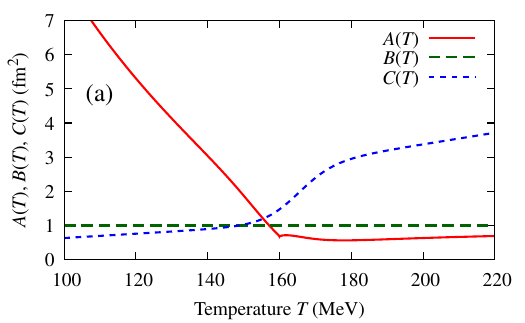}\\
  \includegraphics[width=0.45\textwidth]{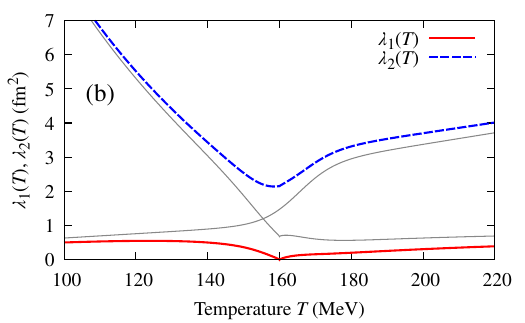}
  \caption{(a) Temperature dependence of Ginzburg-Landau parameters $A(T),
  B(T)$, and $C(T)$.  The solid red line, the green long-dashed line, and
  the blue dashed line show $A(T)$, $B(T)$, and $C(T)$, respectively.  (b)
  Temperature dependence of eigenvalues of the coefficient matrix of the
  potential $V(n, \sigma)$~\eqref{eq:GL_V}.}
  \label{Fig:parameter_A}
\end{figure}

\begin{figure}[htbp]
  \centering
  \includegraphics[width=0.40\textwidth]{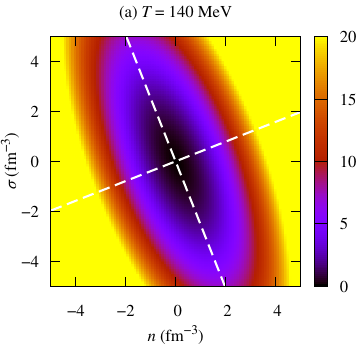}\\
  \includegraphics[width=0.40\textwidth]{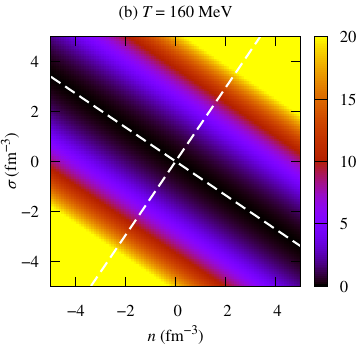}\\
  \includegraphics[width=0.40\textwidth]{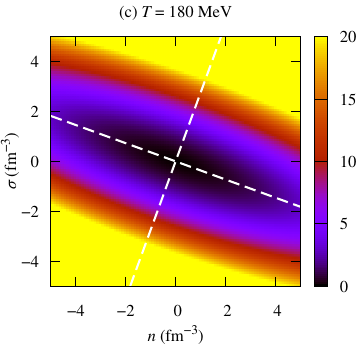}
  \caption{The potential $V(n, \sigma)$ in the Ginzburg-Landau free energy
  in (a) the QGP phase at $T = 180\MeV$,
  (b) the critical temperature at $\Tc = 160\MeV$,
  and (c) the hadron phase at $T = 140\MeV$\@.  The directions of the normal
  modes are indicated by white dashed lines.}
  \label{fig:potential}
\end{figure}

Figure~\ref{Fig:parameter_A} (a) shows the temperature dependence of
the Ginzburg-Landau parameters $A(T)$, $B(T)$, and $C(T)$.  In this model, the
critical behavior appears in $A(T)$; $A(T)$ takes a local minimum at the
critical temperature $\Tc = 160\MeV$.  In contrast, $C(T)$ has smooth
temperature dependence because $C(T) = T / \chi^\mathrm{reg}(T)$ is determined
by the regular part as in Eq.~\eqref{eq.param.C}.
Figure~\ref{Fig:parameter_A} (b) shows the eigenvalues of the
coefficient matrix of the potential $V(n, \sigma)$~\eqref{eq:GL_V}, which
correspond to the curvatures of the potential in the directions of its
principal axes.  The gray thin lines show $A(T)$ and $C(T)$ [which are the same
as the red and blue lines in Fig.~\ref{Fig:parameter_A} (a)] and correspond to the eigenvalues
in the no-mixing case, $B=0$.  We see that the coupling $B$ between $n$ and
$\sigma$ in the potential causes level splitting at the crossing point of the
two gray lines and pushes down and up $\lambda_1(T)$ and $\lambda_2(T)$,
respectively.  As a result, $\lambda_1(T)$ vanishes at the critical temperature
$\Tc = 160\MeV$, which means that the flat direction appears in the potential
landscape.

Figure~\ref{fig:potential} shows the potential shapes of $V(n, \sigma)$ in the
Ginzburg-Landau free energy $F$.  The white dashed lines show the potential axis
directions which correspond to the eigenvectors of the coefficient matrix of $V(n, \sigma)$ associated with the eigenvalues
$\lambda_1$ and $\lambda_2$.  We observe the flat direction at the critical
temperature $\Tc = 160\MeV$\@.  We also see the rotation of the potential-axis
directions in different temperatures.

\begin{figure}[htbp]
  \includegraphics[width=0.45\textwidth]{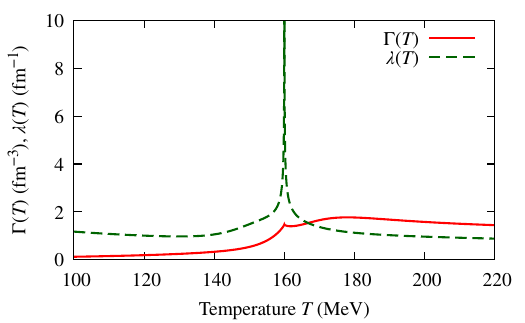}
  \caption{Temperature dependence of the kinetic coefficients $\Gamma(T)$ and
  $\lambda(T)$.  The solid red line and the green dashed line show $\Gamma(T)$
  and $\lambda(T)$, respectively.  }
  \label{Fig:parameter_gamma}
\end{figure}
Figure~\ref{Fig:parameter_gamma} shows the temperature dependence of the
kinetic coefficients $\Gamma(T)$ and $\lambda(T)$.  We omitted
$\tilde\lambda(T)$ in Fig.~\ref{Fig:parameter_gamma} because we set $\tilde\lambda(T) = 0$.  In
this setup, $\Gamma(T)$ and $\lambda(T)$ both increase at the critical
temperature $\Tc = 160\MeV$.
In our parametrization, $\lambda(T)$ diverges at the critical point.
This results from a non-linear coupling of the enhanced fluctuations, whose explicit treatment is beyond the present linearized framework.

The baryon relaxation time $\tauR$ is assumed to be a constant without
temperature dependence, for which we consider several values, $\tauR = 0$, 2.0,
4.0\fm.
It should be noted that with a constant $\tauR$ ($\ne 0$), the signal
propagation speed $v^2$~\eqref{eq:dispersion.3x3-highk-prop} becomes larger
than unity near the critical point in a short period, which can be physically a
problem.  For the present qualitative analysis, we stick to the constant baryon
relaxation times.  However, the temperature-dependent relaxation time needs to
be considered in the future realistic simulation of heavy-ion collision
reactions.

\section{Green's functions}
\label{sec:Green}
We investigate the behavior of the (1+1)d $n$--$\sigma$--$\nu$ system in
expanding background, which we introduced in
Sec.~\ref{sec:model.n-sigma-nu.expanding}\@.  The detailed numerical
implementation is given in Appendix~\ref{app:numerical}.

\begin{figure}[hbt]
  \centering
  \includegraphics[width = 0.48\textwidth]{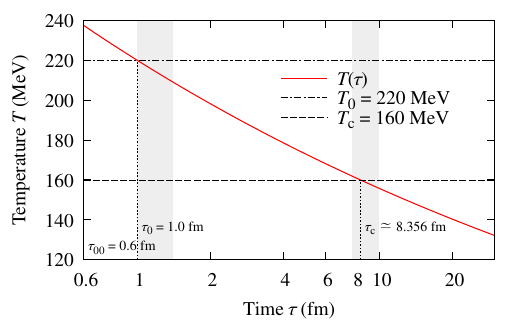}
  \caption{Temperature as a function of the time $\tau$ shown in the red solid
  line.  The black dot-dashed and dashed lines show the initial and critical
  temperature, $T_0$ and $\Tc$, respectively.  The vertical dot lines show the
  initial time, $\tau_0$, and the time the system passes the critical
  point, $\tau_\mathrm{c}$.  The shades show the observation durations of the
  Green's functions in Sec.~\ref{sec:Green}.  The left edge corresponds to
  $\tau_{00}=0.6\fm$, which will be used in Sec.~\ref{sec:Corr}.}
  \label{fig:param.temperature}
\end{figure}

Since the $n$--$\sigma$--$\nu$ system is linear differential equations, it is
useful to check the Green's functions of this system, which is the system's
response to the impulse of $\tilde n$, $\sigma$, or $\nu$ at a specific
spacetime position $(\tau, \etas)$.  The set of the Green's functions forms the
fundamental building blocks of the solutions in the linear systems and
represents the characteristics of the dynamics of the system.  Since the
coefficients in
Eqs.~\eqref{eq:critical_fluctuations_eta_tau1}--\eqref{eq:critical_fluctuations_eta_tau3}
is time-dependent, the formal solution is written by time-ordered exponentials.
Instead of numerically solving the time-ordered exponential for the Green's
functions, we here directly obtain numerical solutions by turning off the noise
terms.  In the following discussions, an impulse in the initial condition will
be given by the Gaussian profile in $\etas$ with the width $w=0.02$ as a
smeared delta function.

We consider two cases for the impulse time, $\tauI = 1.0$ and $7.7\fm$, and
observe the behavior until $\tau = 1.4$ and $10.0\fm$, respectively.  The two
observation durations are shown by shades in Fig.~\ref{fig:param.temperature}.
Figure~\ref{fig:param.temperature} shows the background temperature as a
function of time $\tau$.  The temperatures at the impulse times, $\tauI = 1.0$
and $7.7\fm$, are $T(\tauI) = 220\MeV$ and $T(\tauI) \simeq 161.98\MeV$,
respectively.  The temperature crosses the critical temperature $\Tc = 160\MeV$
at $\tau = \tau_\mathrm{c} \simeq 8.356\fm$.

We first see the profiles of the Green's functions in the $n$--$\sigma$--$\nu$
system (and the $n$--$\sigma$ system as the vanishing $\tauR$ case) in response
to the impulses in $\tilde n$ and $\sigma$ in Secs.~\ref{sec:green.n-impulse}
and~\ref{sec:green.sigma-impulse}, respectively.  For comparison, we also check
the response of the $n$--$\nu$ system (and the $n$ system as the $\tauR=0\fm$
case) to the $\tilde n$-impulse in Sec.~\ref{sec:green.n-nu-system}.  For the
full set of the Green's functions, we also need to consider the responses to
the impulse in the diffusion current $\nu$, which are given for both the
$n$--$\sigma$--$\nu$ and $n$--$\nu$ systems in
Appendix~\ref{app:green.nu-impulse}\@.  In Sec.~\ref{sec:green.spacetime}, we
observe the propagation of waves in the $n$--$\sigma$-$\nu$ system with
$\tauR=4.0\fm$ in the spacetime diagrams.

\begin{figure*}[htbp]
  \centering
  \includegraphics[width=0.99\textwidth]{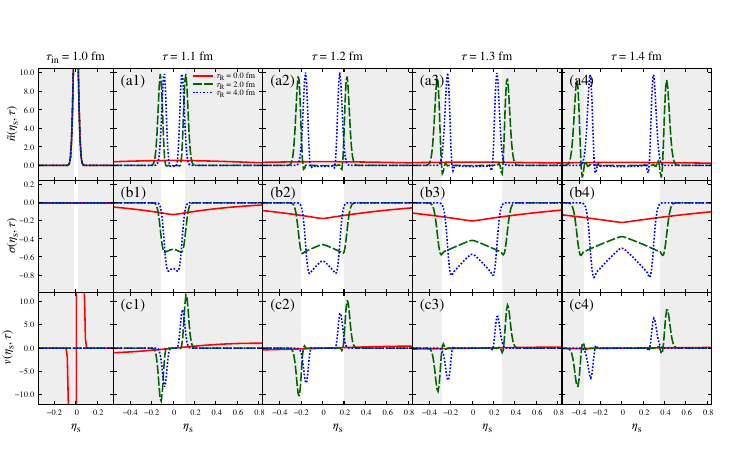}
  \includegraphics[width=0.99\textwidth]{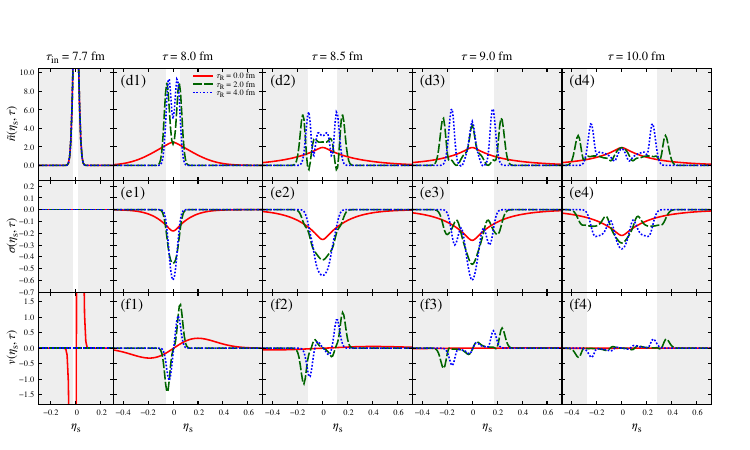}
  \caption{Spacetime evolution of $\tilde n(\etas, \tau)$, $\sigma(\etas,
  \tau)$, and $\nu(\etas, \tau)$ in response to the initial impulse in $\tilde
  n$ in the $n$--$\sigma$--$\nu$ system.  The
  red solid line, the green dashed line, and the blue dotted line are for
  vanishing relaxation time, $\tauR=2.0\fm$, and $\tauR=4.0\fm$, respectively.
  Rows (a)--(c) show the evolution from $\tauI = 1.0\fm$ at
  $\tau = 1.1$, 1.2, 1.3, and 1.4\fm.  Rows (d)--(f) show the evolution from
  $\tauI = 7.7\fm$ at $\tau = 8.0$, 8.5, 9.0, and 10.0\fm.  The
  shade represents the acausal region outside the light cone.}
  \label{Fig:evolution_baryon}
\end{figure*}

\begin{figure*}[htbp]
  \centering
  \includegraphics[width=0.99\textwidth]{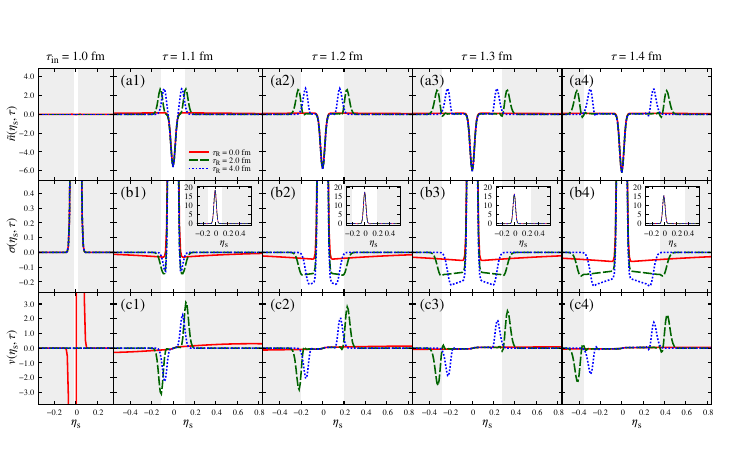}
  \includegraphics[width=0.99\textwidth]{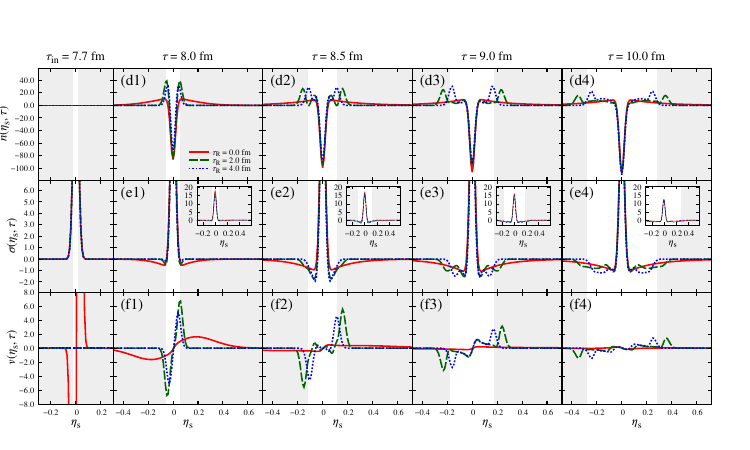}
  \caption{Same as Fig.~\ref{Fig:evolution_baryon} but in response to the
  initial impulse in $\sigma$ instead of $\tilde n$.
  }
  \label{Fig:evolution_sigma}
\end{figure*}

\begin{figure*}[htbp]
  \centering
  \includegraphics[width=0.99\textwidth]{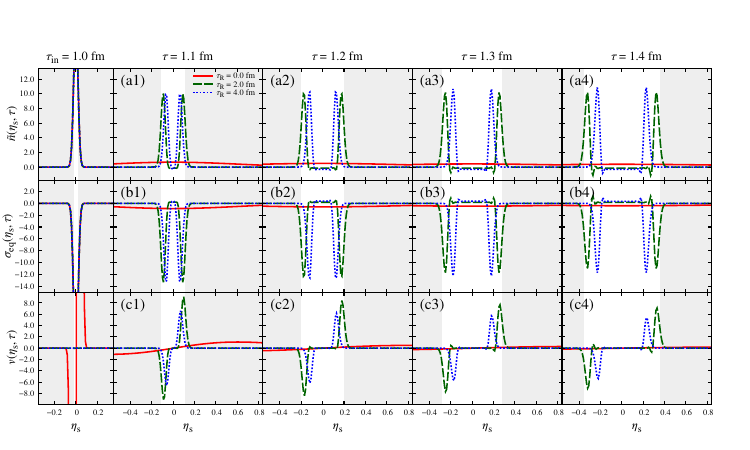}
  \includegraphics[width=0.99\textwidth]{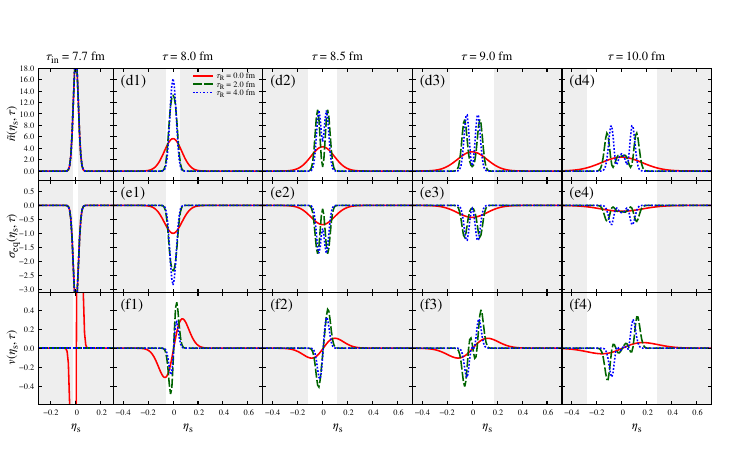}
  \caption{Same as Fig.~\ref{Fig:evolution_baryon} but in the $n$--$\nu$ system
  instead of the $n$--$\sigma$--$\nu$ system.  The impulse is injected in
  $\tilde n(0, \tauI)$.  In this system, $\sigma$ is not dynamical but
  determined by $\sigma_\mathrm{eq} = -(B/A)(\tilde n/\tau)$.
  }
  \label{Fig:evolution.n-nu.n}
\end{figure*}

\subsection{Response to \texorpdfstring{$\tilde n$}{ñ}-impulse}
\label{sec:green.n-impulse}
Figure~\ref{Fig:evolution_baryon} shows the spacetime evolution in response to
the $\tilde n$-impulse at $\etas = 0$ and $\tau = \tauI$:
\begin{align}
  \tilde n(\etas, \tauI)
  &= \frac{1}{\sqrt{2\pi w^2}} e^{-\etas^2/2w^2}\;,
  \label{eq:green.impulse.n}
\end{align}
with $w = 0.02$.  The other fields are initialized to vanish: $\sigma(\etas,
\tauI) = 0$ with $\tauR=0\fm$, and $\sigma(\etas, \tauI) = \nu(\etas, \tauI) =
0$ with $\tauR>0\fm$.  The upper three rows (a)--(c) of
Fig.~\ref{Fig:evolution_baryon} show the time evolutions of $\tilde n(\etas,
\tau)$, $\sigma(\etas, \tau)$, and $\nu(\etas, \tau)$, respectively, for the
impulse time $\tauI = 1.0\fm$.  The lower three rows (d)--(f) of
Fig.~\ref{Fig:evolution_baryon} show the results for the impulse time $\tauI =
7.7\fm$.  The shade shows the acausal region outside the light cone, defined by
$(t - \tauI)^2 - z^2 = 0$, with the $\etas$-margin $w$.

The first row (a) of Fig.~\ref{Fig:evolution_baryon} shows the baryon density
fluctuation $\tilde n$ for different baryon relaxation times $\tauR = 0.0$,
$2.0$, and $4.0\fm$.  We notice that the initial peak of the vanishing $\tauR$
case (red line) in (a1) at $\tau=1.0\fm$ disappears in (a2) at
$\tau=1.1\fm$, while the peak splits into two propagating peaks in the other
$\tauR$ cases (blue and green lines).  With the vanishing relaxation time,
$\tilde n$ instantly becomes finite outside the light cone, which violates
causality.  With the finite baryon relaxation time, while being diffused
slowly, the two peaks move to the $\pm\etas$ directions in a finite speed $v$
determined by Eq.~\eqref{eq:dispersion.3x3-highk-prop}:
\begin{align}
  v \simeq \biggl(\frac{3.27\fm}{\tauR}\biggr)^{1/2}\;, \quad (T = 220\MeV)\;.
\label{eq:green.velocity}
\end{align}
We observe that the peaks remain in the causal region with $\tauR=4.0\fm$ where
$v<1$, while the peaks intrude into the acausal region with $\tauR=2.0\fm$
where $v>1$. We see similar trends in the propagation speed of the structure
in the other time evolutions shown in (b)--(f).

The second row (b) of Fig.~\ref{Fig:evolution_baryon} shows the time evolutions
of the $\sigma$ fluctuations.  The negative $\sigma$ is induced by the positive
$n$ through the $\sigma$--$n$ coupling in
Eq.~\eqref{eq:critical_fluctuations_eta_tau2}; the overall coupling $-\Gamma
B/\tau$ is negative since both coefficients $B(T)$ and $\Gamma(T)$ (shown in
Figs.~\ref{Fig:parameter_A} and \ref{Fig:parameter_gamma}, respectively) are
positive.

The third row (c) of Fig.~\ref{Fig:evolution_baryon} shows the time evolution
of the diffusion current $\nu$. In the diffusion current $\nu$, positive and
negative peaks travels to the $\pm\etas$ directions, respectively.  It should
be noted that, in the case of $\tauR=0$, we evaluate $\nu$ by the right-hand
side of Eq.~\eqref{eq:critical_fluctuations_eta_tau3} although $\nu$ is not
dynamical.

The lower three rows (d)--(f) of Fig.~\ref{Fig:evolution_baryon} are similar to
(a)--(c) but with the impulse time $\tauI= 7.7\fm$.  In this setup, we
observe the behavior of the solution when the system crosses the critical
temperature $\Tc = 160\MeV$ at $\tau \simeq 8.356\fm$: panels (d1), (e1), and
(f1) of Fig.~\ref{Fig:evolution_baryon} at $\tau = 8.0\fm$ show the profiles
before the temperature of the system crosses $\Tc$, while panels (d2), (e2), and (f2) of
Fig.~\ref{Fig:evolution_baryon} at $\tau = 8.5\fm$ show those after the temperature of the system
goes below $\Tc$.  We notice that with finite baryon relaxation times, new bumps
are created near $\etas\sim0$ in panel (d2), after the system undergoes $\Tc$,
and then merge with each other in (d3) and (d4) of
Fig.~\ref{Fig:evolution_baryon}.  This new structure in $\tilde n$ creates the
structure in $\sigma$ and $\nu$ as well as seen in (e1)--(f4) of
Fig.~\ref{Fig:evolution_baryon}.

\subsection{Response to \texorpdfstring{$\sigma$}{σ}-impulse}
\label{sec:green.sigma-impulse}
In Fig.~\ref{Fig:evolution_sigma}, we also check the responses to the
$\sigma$-impulse at $\tau = \tauI$:
\begin{align}
  \sigma(\etas, \tauI) &= \frac1{\sqrt{2\pi w^2}} e^{-\etas^2/2w^2}\;,
  \label{eq:green.impulse.sigma}
\end{align}
with $w = 0.02$ and the other fields being $\tilde n (\etas, \tauI) =
\nu(\etas, \tauI) = 0$.  Similarly to Fig.~\ref{Fig:evolution_baryon}, the
upper three rows (a)--(c) and the lower three rows (d)--(f) of
Fig.~\ref{Fig:evolution_sigma} show the results with the impulse time $\tauI=
1.0\fm$ and 7.7\fm, respectively.

In the first row (a) of Fig.~\ref{Fig:evolution_sigma}, we see that negative
$\tilde n$ is induced near the origin $\etas = 0$ for all the cases of the
baryon relaxation time $\tauR$.  This is caused by
$\partial^2\sigma/\partial\etas^2 < 0$ near the origin.  By combining
Eqs.~\eqref{eq:critical_fluctuations_eta_tau1}
and~\eqref{eq:critical_fluctuations_eta_tau3} and neglecting $\tauR$, we obtain
\begin{align}
  \frac{\partial\tilde n}{\partial\tau}
  &\sim \frac{\tilde \lambda A + \lambda B}{\tau}\frac{\partial^2 \sigma}{\partial\etas^2}\;,
\end{align}
where the coefficient $(\tilde\lambda A + \lambda B)/\tau$ is positive because
$\tilde\lambda = 0$, and $B(T)$ and $\lambda(T)$ (shown in
Figs.~\ref{Fig:parameter_A} and~\ref{Fig:parameter_gamma}, respectively) are
always positive.  Because of the initial peak in $\sigma$,
$\partial^2\sigma/\partial\etas^2$ is negative around $\etas = 0$ and thus decreases
the baryon density $\tilde n$.  Due to the charge conservation, the decrease
needs to be compensated by the increase in other regions.  We here see a clear
behavioral difference between the finite and vanishing baryon relaxation times:
clear positive bumps are created beside the negative peak when $\tauR$ is
finite, while they do not appear with the vanishing $\tauR$.  Instead, baryon
density is slightly enhanced at a wide range of $\etas$.

In the second row (b) of Fig.~\ref{Fig:evolution_sigma}, we see that negative
dips of $\sigma$ beside the initial peak are induced by the positive $\tilde
n$.  In the case of finite $\tauR$, as the positive $\tilde n$ peaks travel to
the $\pm\etas$ directions, the $\sigma$ dips are expanded to form a potential
well-like structure.  In the case of vanishing $\tauR$, the dips have smooth
and shallow Gaussian tails in a wide range of $\etas$ from the beginning and do
not change their shapes.  Meanwhile, the central peak of $\sigma$ decreases
over time as seen in the insets of Fig.~\ref{Fig:evolution_sigma} (b).

The evolutions of $\tilde n$ and $\sigma$ from $\tauI=7.7\fm$ are shown in the
fourth and fifth rows, (d) and (e), of Fig.~\ref{Fig:evolution_sigma}.  The
qualitative behavior seems similar to the case of $\tauI=1.0\fm$, but
additional structures appear when the system passes the critical temperature
$\Tc$.  As a result, a clear potential well-like structure is not seen in this
case.

The evolution of the diffusion currents $\nu$ in the third and sixth rows, (c)
and (f), of Fig.~\ref{Fig:evolution_sigma} is found to be the same as those
for the $\tilde n$-impulse in Fig.~\ref{Fig:evolution_baryon} (c) and (f).
This is because the dynamics of $\nu$ is governed by the linear combination of
$\sigma$ and $\tilde n$ in Eq.~\eqref{eq:critical_fluctuations_eta_tau3}, where
the difference of the $\sigma$-impulse and the $\tilde n$-impulse is
insignificant concerning the dynamics of $\nu$.

\subsection{Response to \texorpdfstring{$\tilde n$}{ñ}-impulse in \texorpdfstring{$n$--$\nu$}{n--ν} system}
\label{sec:green.n-nu-system}
We can compare the Green's functions of the $n$--$\sigma$-$\nu$ system shown in
previous subsections with the $n$--$\sigma$ system, where $\sigma$ is replaced
by its equilibrium value $\sigma_\mathrm{eq} = -(B/A)(\tilde n/\tau)$.
Figure~\ref{Fig:evolution.n-nu.n} shows the time evolution of the $n$--$\nu$
system in response to the $\tilde n$-impulse~\eqref{eq:green.impulse.n}.  In
the second and fifth rows, the equilibrium value $\sigma_\mathrm{eq}$ is shown
instead of the dynamical $\sigma$.

The qualitative behaviors of the dynamical degrees of freedom, $\tilde n$ and
$\nu$, in Fig.~\ref{Fig:evolution.n-nu.n} are similar to those of the
$n$--$\sigma$--$\nu$ system in Fig.~\ref{Fig:evolution_baryon}, but their
structures are sharper in the $n$--$\nu$ system.  We also notice that the
propagation speeds of the waves are smaller in the $n$--$\nu$ system
than in the $n$--$\sigma$--$\nu$ system,
which is
because the underlying dispersion relations,
Eqs.~\eqref{eq:dispersion.3x3-highk-prop}
and~\eqref{eq:dispersion.n-nu.general}, are different in the
$n$--$\sigma$--$\nu$ and $n$--$\nu$ systems.  In the current setup with $\tilde
\lambda = 0$ and Eq.~\eqref{eq.param.C}, the signal propagation speed in the
$n$--$\sigma$--$\nu$ system becomes
\begin{align}
  v_{n\text{--}\sigma\text{--}\nu} = \sqrt{\frac{\tilde \lambda B + \lambda C}{\tauR}}
  = \sqrt{\frac{\lambda T}{\tauR\chi^\mathrm{reg}}}\;.
\end{align}
The signal propagation speed in the $n$--$\nu$ system is obtained as
\begin{align}
  v_{n\text{--}\nu} = \sqrt{\frac{D}{\tauR}}
  = \sqrt{\frac{\lambda T}{\tauR\chiB}}\;,
\end{align}
using Eq.~\eqref{eq:param.D}. Since $\chiB = \chi^\mathrm{cr} +
\chi^\mathrm{reg} > \chi^\mathrm{reg}$, the signal propagation speed in the
$n$--$\nu$ system becomes smaller: $v_{n\text{--}\nu} <
v_{n\text{--}\sigma\text{--}\nu}$.

We may also compare the equilibrium values $\sigma_\mathrm{eq}$ in
Fig.~\ref{Fig:evolution.n-nu.n} with the dynamical values of $\sigma$ in
Fig.~\ref{Fig:evolution_baryon}.  In the $n$--$\nu$ system where $\sigma$
instantaneously relaxes to the equilibrium value $\sigma_\mathrm{eq}$, the
magnitude of $\sigma$ is significantly larger than in the $n$--$\sigma$--$\nu$
system. This implies that the timescale in which the peaks pass through a
particular region is significantly shorter than the relaxation time of
$\sigma$, $\tau_\sigma = 1/\Gamma A \sim 1\fm$.

\subsection{Spacetime diagrams}
\label{sec:green.spacetime}
\begin{figure*}[htbp]
  \centering
  \includegraphics[width=0.9\textwidth]{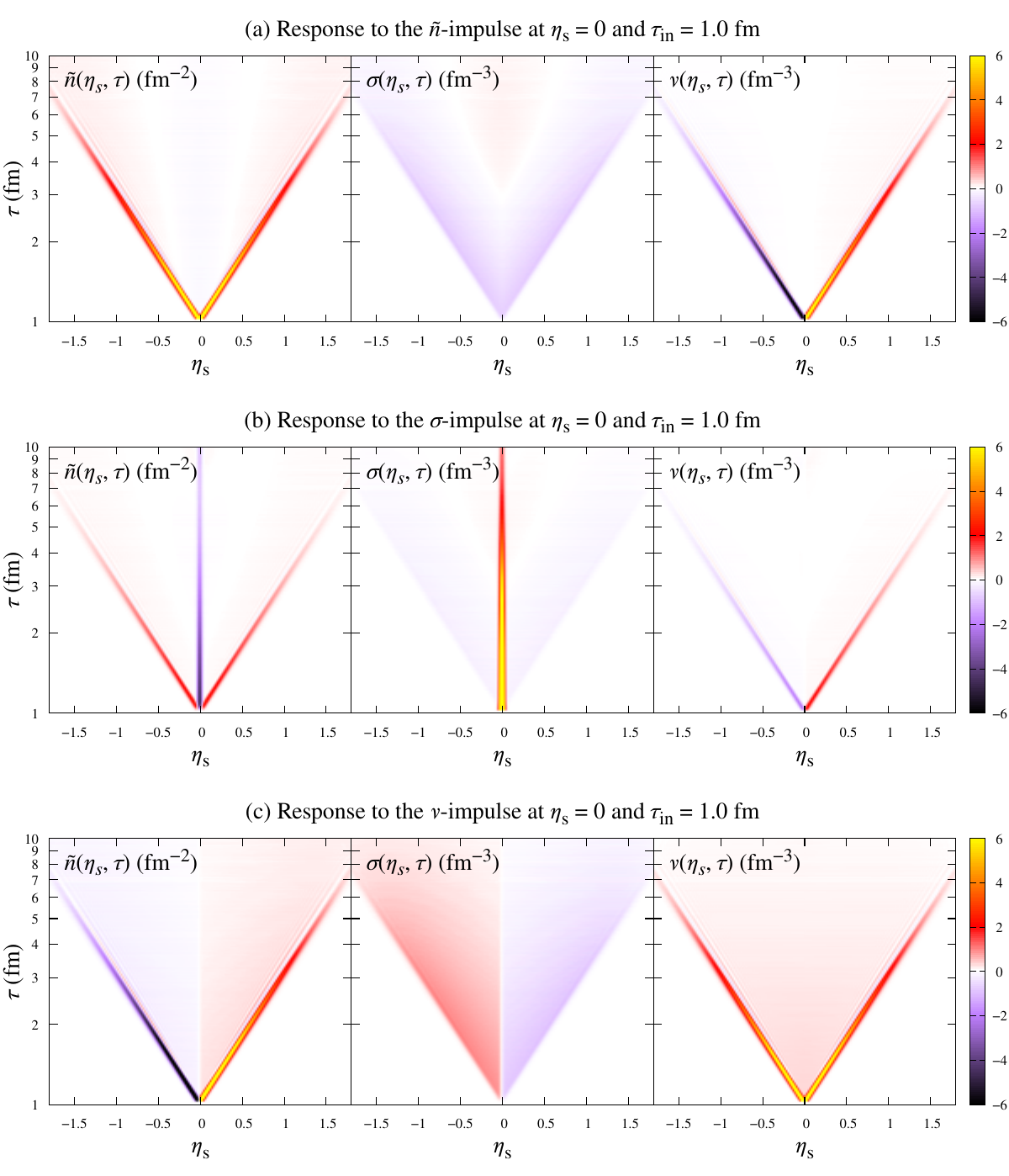}
  \caption{Spacetime diagrams of $\tilde n(\etas, \tau)$, $\sigma(\etas,
  \tau)$, and $\nu(\etas, \tau)$ in response to impulses at $\tauI =
  1.0\fm$ with the relaxation time $\tauR = 4.0\fm$.  Rows (a), (b), and
  (c) correspond to the impulses at $\tilde n(0,
  \tauI)$~\eqref{eq:green.impulse.n}, $\sigma(0,
  \tauI)$~\eqref{eq:green.impulse.sigma}, and $\nu(0,
  \tauI)$~\eqref{eq:green.impulse.nu}, respectively.  These
  structures in (a), (b), and (c) correspond to blue lines in the upper three
  rows of Figs.~\ref{Fig:evolution_baryon}, \ref{Fig:evolution_sigma}, and
  \ref{Fig:evolution_nu}, respectively.}
  \label{Fig:evolution_3d_1.0}
\end{figure*}
\begin{figure*}[htbp]
  \centering
  \includegraphics[width=0.9\textwidth]{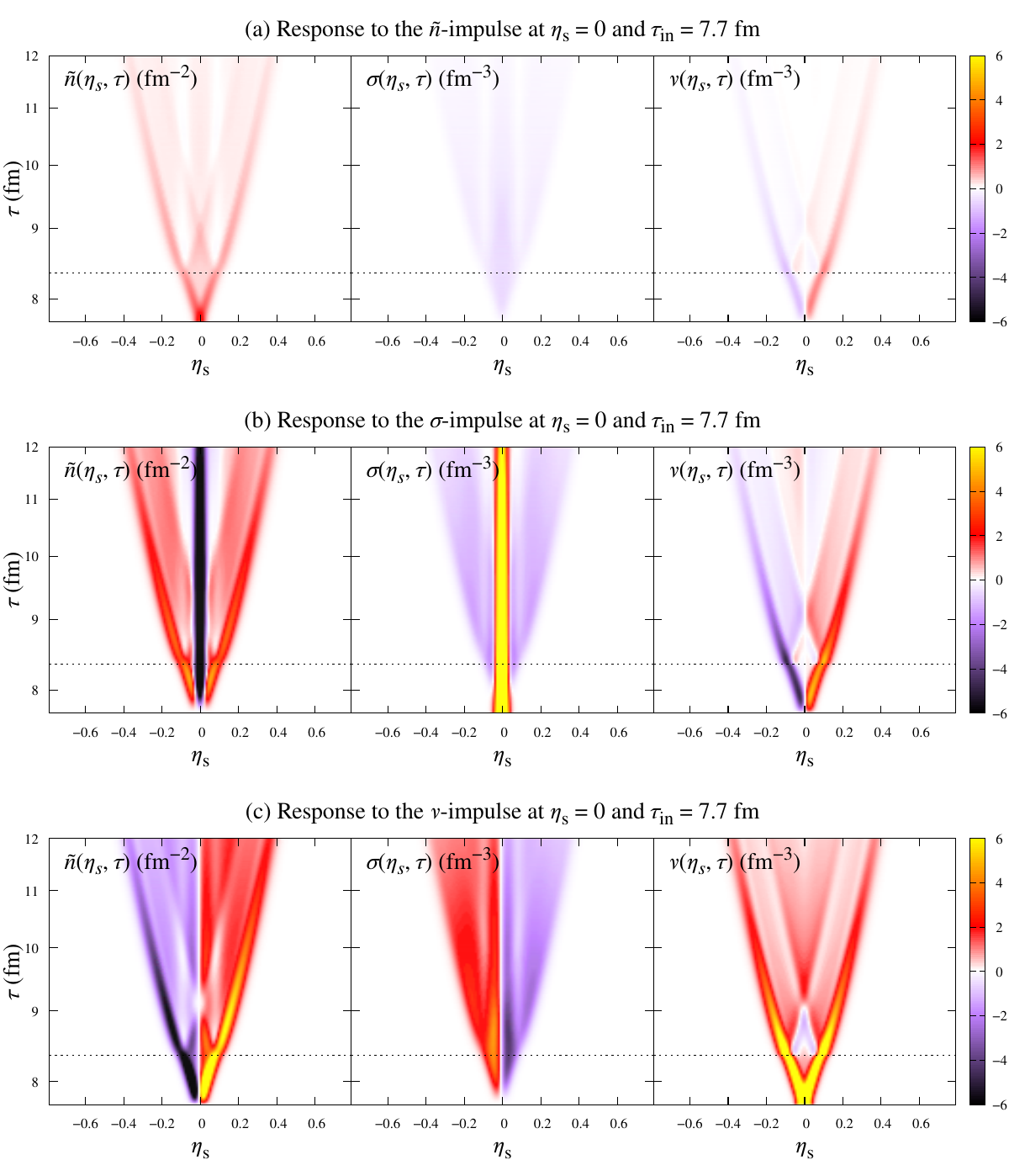}
  \caption{Spacetime diagrams of $\tilde n(\etas, \tau)$, $\sigma(\etas,
  \tau)$, and $\nu(\etas, \tau)$ similar to Fig.~\ref{Fig:evolution_3d_1.0},
  but for the impulses at $\tauI = 7.7\fm$.  Rows (a), (b), and (c)
  correspond to blue lines in the lower three rows of
  Figs.~\ref{Fig:evolution_baryon}, \ref{Fig:evolution_sigma}, and
  \ref{Fig:evolution_nu}, respectively.  The horizontal dashed line indicates
  the time the system goes through $\Tc$.}
  \label{Fig:evolution_3d_7.7}
\end{figure*}
It is also useful to see the time evolutions of fields in spacetime diagrams.
We here stick with the causal value of the baryon relaxation time, $\tauR =
4.0\fm$, as discussed using Eq.~\eqref{eq:green.velocity}.
Figure~\ref{Fig:evolution_3d_1.0} shows the spacetime evolutions of $\tilde
n(\etas, \tau)$, $\sigma(\etas, \tau)$, and $\nu(\etas, \tau)$ for impulses at
$\tauI = 1.0\fm$.  Panels (a), (b), and (c) show the evolutions after the
impulses in $\tilde n(0, \tauI)$, $\sigma(0, \tauI)$, and $\nu(0, \tauI)$,
respectively.  The vertical axis is chosen to be $\ln\tau$ so that the slope
reflects the wave propagation speed on the Bjorken background\footnote{When a
wave propagates at the velocity $v$ inside the one-dimensionally expanding
medium, the wave equation is written by $[\partial/\partial t + v
(1/\tau)\partial/\partial\etas]\phi = 0$.  This is equivalent to
$(\partial/\partial\ln\tau + v\partial/\partial\etas)\phi = 0$, which can be
regarded as the equation of the right-moving wave at the velocity $v$ in the
$(\ln\tau)$--$\etas$ space.}.

In all the cases, we see the wave fronts move at a constant speed in the right
and left directions.  In Fig.~\ref{Fig:evolution_3d_1.0} (b), the central peak
at $\etas = 0$ remains in $\tilde n$ and $\sigma$ for a long time while
decreasing gradually.

Figure~\ref{Fig:evolution_3d_7.7} shows the spacetime diagrams when the
impulses are injected at $\tauI = 7.7\fm$. The horizontal dashed line shows the
time when the system crosses the critical temperature $\Tc = 160\MeV$ at $\tau
\simeq 8.356\fm$.  We observe each peak split into two peaks at $\Tc$ and
propagate to $\pm\etas$ as if the reflected and transmitted waves.
This is due to
the rapid change of the linear combination of the propagating
modes~\eqref{eq:dispersion.3x3-highk-prop} near the critical point,
which is caused by the divergence of the kinetic coefficient $\lambda$ within our model.
The
reflected waves toward the origin cross each other at $\tau \sim 9.0\fm$.  The
peak at $\etas = 0$ seen in Fig.~\ref{Fig:evolution_baryon}~(d3) can be
understood by the superposition of the reflected waves created at $\Tc$.

\section{Two-point correlations}
\label{sec:Corr}

We analyze the equal-time two-point correlation functions in dynamical systems
with noise terms $\xi_\sigma$ and $\xi_\nu$.  We first consider the
$n$--$\sigma$--$\nu$
system~\eqref{eq:critical_fluctuations_eta_tau1}--\eqref{eq:critical_fluctuations_eta_tau3}
in Sec.~\ref{subsec:n-sigma-nu} and then the $n$--$\nu$
system~\eqref{eq:model.n-nu-expanding.eom1}--\eqref{eq:model.n-nu-expanding.eom2}
in Sec.~\ref{subsec:n-nu}\@.  We also check the temperature dependence of the
ultraviolet peak in the correlation in Sec.~\ref{sec:result.2pt.peak-vs-T}.

We perform simulations of 4\ 480 events for each setup. The calculation is
performed in a one-dimensional box of the size $L=12$ with the periodic
boundary condition, $\phi_i(\etas+L, \tau) = \phi_i(\etas, \tau)$, where
$\bm{\phi} = (\phi_1, \phi_2, \phi_3) := (\tilde n, \sigma, \nu)$.  To obtain
the initial conditions at $\tau_0=1.0\fm$, we set the vanishing fields, $\tilde
n(\etas, \tau_{00}) = \sigma(\etas, \tau_{00}) = \nu(\etas, \tau_{00}) = 0$, at
$\tau_{00} = 0.6\fm$ and solve the time evolution until $\tau_0 = 1.0\fm$ to
build up a stable distribution of an expanding system.  We note that the
initialization by the thermal distribution at $\tau_0$ turned out to generate
spurious waves in the correlations because the thermal distribution in the
static equilibrium is not necessarily a stable distribution in the expanding
system.

We start observation at $\tau_0 = 1.0\fm$.  We calculate the two-point
correlations at an equal proper time $\tau \;(\ge\tau_0)$ by spatial and event
averages:
\begin{multline}
  \langle\phi_i(\Delta\etas, \tau)\phi_j(0, \tau)\rangle \\
    := \frac1{N_\mathrm{ev}} \sum_{k = 1}^{N_\mathrm{ev}}
    \int_0^L \frac{d\etas}{L} \phi_i^{(k)}(\etas + \Delta\etas, \tau)\phi_j^{(k)}(\etas, \tau)\;,
\end{multline}
where $\phi_i^{(k)}(\etas, \tau)$ is the solution of the $k$th event.

\subsection{\texorpdfstring{$n$--$\sigma$--$\nu$}{n--σ--ν} and \texorpdfstring{$n$--$\sigma$}{n--σ} systems}
\label{subsec:n-sigma-nu}
\begin{figure*}[htbp]
  \includegraphics[width=0.9\textwidth,  bb=20 15 360 140]{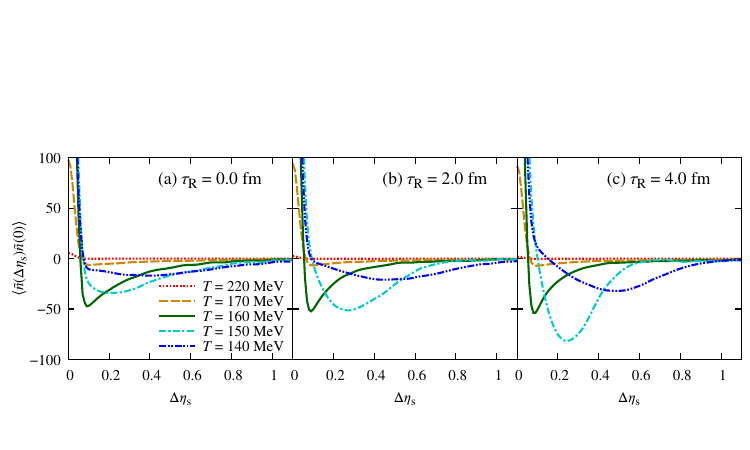}
  \includegraphics[width=0.9\textwidth,  bb=20 15 360 140]{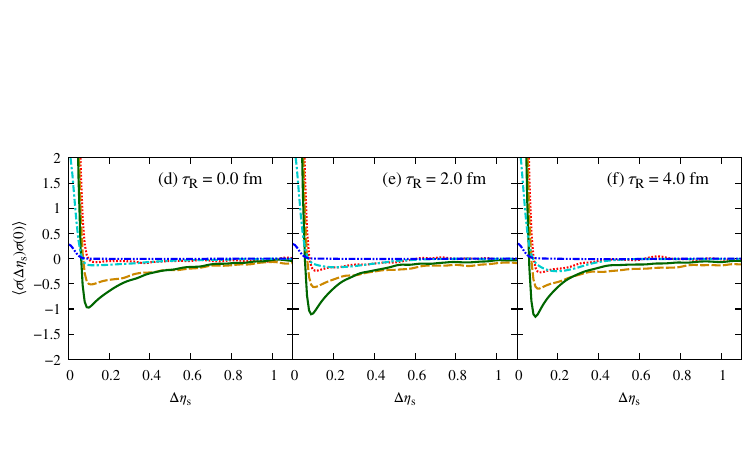}
  \includegraphics[width=0.9\textwidth,  bb=20 15 360 140]{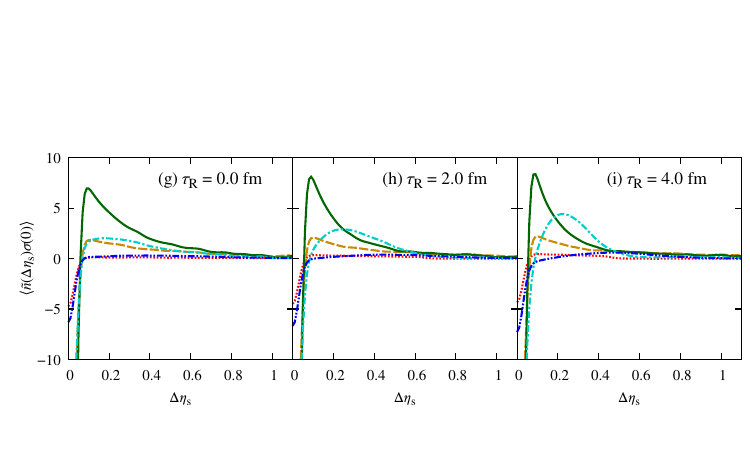}
  \caption{The two-point correlations in the $n$--$\sigma$--$\nu$ system in the
  one-dimensionally expanding system.  The upper row (a)--(c), the middle row
  (d)--(f), and the bottom row (g)--(i) show $\langle \tilde n(\Delta\etas)
  \tilde n(0) \rangle$, $\langle \sigma(\Delta\etas) \sigma(0) \rangle$, and
  $\langle \tilde n(\Delta\etas) \sigma(0) \rangle$, respectively.  The left,
  center, and right columns show the results with different relaxation times,
  $\tauR = 0$, $2.0$, and $4.0\fm$.}
\label{Fig:correlation_baryon_sigma}
\end{figure*}
\begin{figure*}[htbp]
  \includegraphics[width=0.9\textwidth]{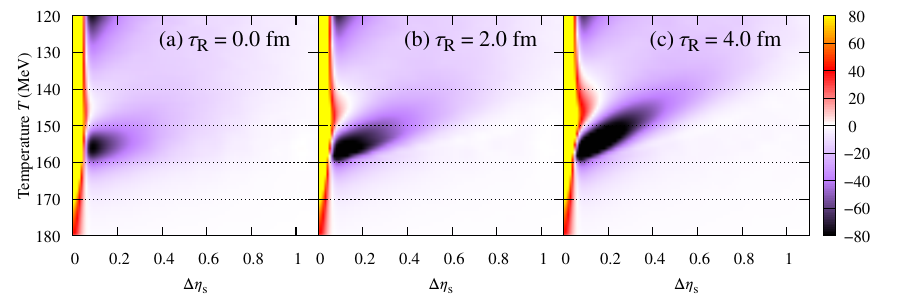}
  \caption{Baryon autocorrelation $\langle\tilde n(\Delta\etas) \tilde
  n(0)\rangle$ as a function of $(\etas, T)$ in the $n$--$\sigma$--$\nu$
  system.  Panel (a) is for vanishing relaxation time, (b) $\tauR = 2.0\fm$,
  and (c) $\tauR = 4.0\fm$.  The dotted lines correspond to the temperatures
  shown in Fig.~\ref{Fig:correlation_baryon_sigma}.}
  \label{Fig:corr3d.n-sigma-nu}
\end{figure*}

We first analyze the $n$--$\sigma$--$\nu$ system, which includes the
$n$--$\sigma$ system as a special case of $\tauR = 0\fm$.
Figure~\ref{Fig:correlation_baryon_sigma} shows the two-point correlation
functions $\langle \tilde n (\Delta\etas) \tilde n (0) \rangle$, $\langle
\sigma(\Delta\etas) \sigma(0) \rangle$, and $\langle \tilde n(\Delta\etas)
\sigma(0) \rangle$, for different baryon relaxation times $\tauR$ and
temperature $T$.
In all the cases, the autocorrelations $\langle\tilde n(\Delta\etas)\tilde
n(0)\rangle$ and $\langle\sigma(\Delta\etas)\sigma(0)\rangle$ take the maximum
at $\Delta \etas = 0$, which is the general behavior of autocorrelations.  In
this subsection, to focus on the other nontrivial part at $\Delta\etas \ne 0$,
we adjusted the range of the vertical axis excluding these peaks.  The
magnitude of the peak will be discussed in Sec.~\ref{sec:result.2pt.peak-vs-T}.

Except for the peak at $\Delta\etas = 0$, $\langle\tilde n(\Delta\etas)\tilde
n(0)\rangle$ and $\langle\sigma(\Delta\etas)\sigma(0)\rangle$ are always
negative.  This can be understood by the charge conservation within the current
setup.  In our setup, $\tilde n$ is defined to be the deviation of the baryon
charge density from its average, and its initial condition is set to $\tilde
n(\etas,\tau_{00})=0$.  Because of the charge conservation and the periodic
boundary condition, the integration of $\tilde n$ over the spatial period $L$
is conserved during the time evolution:
\begin{align}
  \int_0^L d\etas \tilde n(\etas, \tau)
  \equiv \int_0^L d\etas \tilde n(\etas, \tau_{00}) = 0\;.
\end{align}
In particular, this exactly holds on an event-by-event basis without
fluctuations because the noise terms $\xi_\sigma$ and $\xi_\nu$ do not directly
enter the time derivative of $\tilde n$ in
Eq.~\eqref{eq:critical_fluctuations_eta_tau2}.  As a consequence, the
integration of the baryon-density autocorrelation over the whole region should
vanish as
\begin{align}
  \int_0^L d\etas\langle\tilde n(\etas) \tilde n(0)\rangle
    &= \biggl\langle\biggl[\int_0^L d\etas \tilde n(\etas)\biggr]\tilde n(0)\biggr\rangle \equiv 0\;.
\end{align}
This means that the positive peak of $\tilde n$ at $\Delta\etas = 0$ needs to
be compensated by the negative contributions to $\tilde n$ in the other region
of $\Delta\etas \ne 0$. A similar behavior in the $\sigma$ autocorrelation
appears through the strong anti-correlation of $\sigma$ and $\tilde n$
fluctuations observed at $\Delta\etas=0$ in the bottom row of
Fig.~\ref{Fig:correlation_baryon_sigma}.

The magnitudes of the positive and negative peaks before the system
reaches down to the critical temperature [shown by the red dotted and yellow
dashed lines in Fig.~\ref{Fig:correlation_baryon_sigma}~(a)--(f)] are smaller
than those after the system underwent the critical temperature $\Tc$ [shown by
the green solid, cyan dot-dashed, and blue dot-dot-dashed lines in
Fig.~\ref{Fig:correlation_baryon_sigma}~(a)--(f)].  On the critical temperature
$\Tc=160\MeV$ (the green solid line), due to the enhanced critical
fluctuations, the central peak becomes significantly large, while a large
negative pocket is formed beside the peak.  The shapes of the autocorrelations
$\langle\tilde n(\Delta\etas)\tilde n(0)\rangle$ and
$\langle\sigma(\Delta\etas)\sigma(0)\rangle$ are similar on the critical
temperature $\Tc = 160\MeV$ but qualitatively different from each other just
below the critical temperature, $T = 150\MeV$ (the cyan dot-dashed line): the
large negative pocket expands to the positive $\etas$ direction in
$\langle\tilde n(\Delta\etas)\tilde n(0)\rangle$, while the negative part of
$\langle\sigma(\Delta\etas)\sigma(0)\rangle$ gradually attenuates locally.
This behavior difference reflects the nature of the conserved quantity $\tilde
n$ and non-conserved quantity $\sigma$.

The $n$--$\sigma$ system with $\tauR = 0\fm$ in the first column of
Fig.~\ref{Fig:correlation_baryon_sigma} and the $n$--$\sigma$--$\nu$ system
with $\tauR\neq 0\fm$ in the second and third columns of
Fig.~\ref{Fig:correlation_baryon_sigma} are qualitatively similar, but the
negative pocket in $\langle\tilde n(\Delta\etas)\tilde n(0)\rangle$ below the
critical temperature is more significant with larger $\tauR$.  The negative
correlation can immediately diffuse to large $\Delta\etas$ in the case of the
vanishing baryon relaxation time, $\tauR=0\fm$, while the negative contribution
cannot propagate to large $\Delta\etas$ and accumulate beside the central peak
in the case of the finite baryon relaxation time $\tauR> 0\fm$.  The
$\tauR$-dependence of $\langle\sigma(\Delta\etas)\sigma(0)\rangle$ is weak,
which means that the relaxation time of $n$ does not have a direct influence on
$\sigma$.

The $n$--$\sigma$ correlation $\langle\tilde n(\Delta\etas)\sigma(0)\rangle$ is
shown on the bottom row of Fig.~\ref{Fig:correlation_baryon_sigma}. The strong
anti-correlation at $\Delta\etas = 0$ is due to the negative coupling $-\Gamma
B / \tau\;(< 0)$ of $\sigma$ to $\tilde n$ in
Eq.~\eqref{eq:critical_fluctuations_eta_tau2}.  Due to this strong
anti-correlation, the negative pockets seen in the autocorrelations
$\langle\tilde n(\Delta\etas)\tilde n(0)\rangle$ and
$\langle\sigma(\Delta\etas)\sigma(0)\rangle$ in
Fig.~\ref{Fig:correlation_baryon_sigma}~(a)--(f) are reflected in
$\langle\tilde n(\Delta\etas)\sigma(0)\rangle$ upside down.

To better illustrate the behavior of the $n$ autocorrelation, $\langle\tilde
n(\Delta\etas)\tilde n(0)\rangle$, Fig.~\ref{Fig:corr3d.n-sigma-nu} shows the
correlation as a function of $(\etas,T)$ in a heat map for different relaxation
times $\tauR$.  The temperatures shown in
Fig.~\ref{Fig:correlation_baryon_sigma} correspond to the sections on the black
dotted lines in Fig.~\ref{Fig:corr3d.n-sigma-nu}. We can clearly see that the
negative pocket is formed near $\Delta\etas=0$ on the critical temperature with
all the baryon relaxation times.  However, the propagation of the negative
pocket is different with different relaxation times.  The negative pocket does
not propagate to the larger $\Delta\etas$ region in lower temperature
(corresponding to a later time $\tau$) as seen in
Fig.~\ref{Fig:corr3d.n-sigma-nu}~(a), while the pocket propagates to the larger
$\Delta\etas$ region with the finite relaxation times $\tauR = 2.0$ and
$4.0\fm$ as shown in Fig.~\ref{Fig:corr3d.n-sigma-nu}~(b) and~(c).  The slope
of the wave is steeper for the larger $\tauR=4.0\fm$, which reflects the slower
propagation speed with larger $\tauR$.

\subsection{\texorpdfstring{$n$--$\nu$}{n--ν} and \texorpdfstring{$n$}{n} systems}
\label{subsec:n-nu}
\begin{figure*}[htbp]
\includegraphics[width=0.9\textwidth,  bb=20 15 360 140]{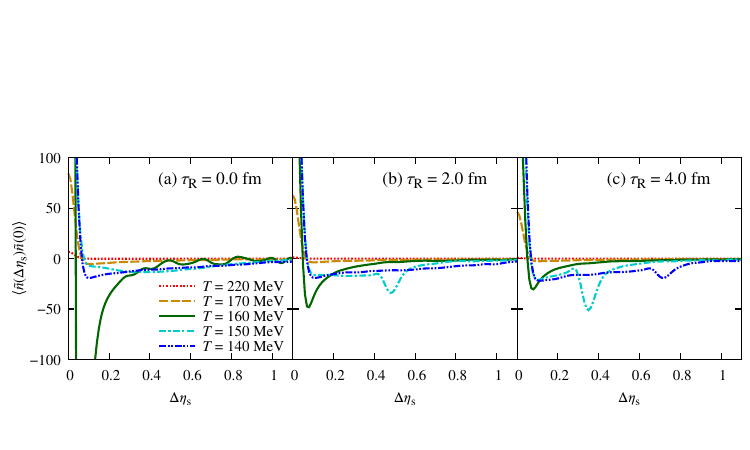}
  \caption{The two-point correlation $\langle \tilde n (0) \tilde n
  (\Delta\etas) \rangle$ for the $n$--$\nu$ system.  Panel (a) is for the
  vanishing relaxation time, (b) $\tauR = 2.0\fm$, and (c) $\tauR = 4.0\fm$.}
  \label{Fig:correlation_nB}
\end{figure*}
\begin{figure*}[htbp]
  \includegraphics[width=0.9\textwidth]{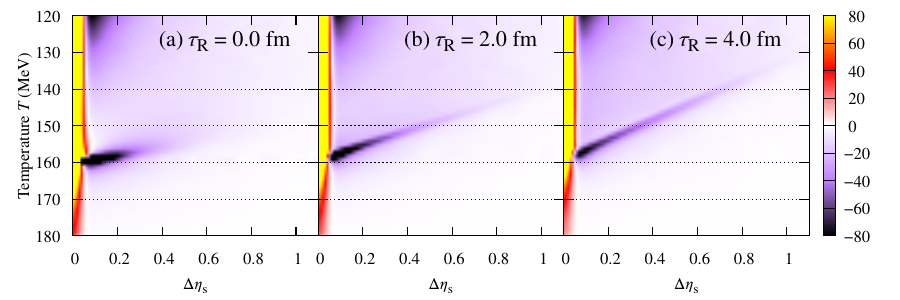}
  \caption{Baryon autocorrelation $\langle\tilde n(\Delta\etas) \tilde
  n(0)\rangle$ as a function of $(\etas, T)$, which is the same as
  Fig.~\ref{Fig:corr3d.n-sigma-nu} but in the $n$--$\nu$ system.}
\label{Fig:correlation_n_3d}
\end{figure*}

For comparison, we also check the two-point correlations in the $n$--$\nu$
system, which is the limiting case of $\tau_\sigma = 1/\Gamma A \to 0$ in the
$n$--$\sigma$--$\nu$ system.  The $n$ system is a special case of the
$n$--$\nu$ system with $\tauR = 0\fm$.

The autocorrelation $\langle \tilde n(\Delta\etas) \tilde n(0) \rangle$ is
shown in Fig.~\ref{Fig:correlation_nB}.  Compared to the $n$--$\sigma$--$\nu$
system shown in Figs.~\ref{Fig:correlation_baryon_sigma}~(a)--(c), the negative
pocket beside the central peak becomes sharper.  In particular, in the
vanishing baryon relaxation case, $\tauR = 0\fm$, the depth of the negative
pocket is significantly large, which also implies the large positive peak.
Conversely, in the case of $\tauR =
4.0\fm$, the negative pocket at the critical temperature $\Tc = 160\MeV$ is
shallower than in the $n$--$\sigma$--$\nu$ system.  These reflect the ordering
of the magnitude of the central peak, which will be discussed in
Sec.~\ref{sec:result.2pt.peak-vs-T}.

We also notice that dips appear at finite $\Delta\etas$ below the critical
temperature for the finite baryon relaxation times $\tauR>0\fm$.  At $T =
150\MeV$, the position of the dip in the case of $\tauR = 4.0\fm$ is closer to
the origin than in the case of $\tauR = 2.0\fm$.  Moreover, the case of $\tauR
= 4.0\fm$ has another dip at $T = 140\MeV$, which is not observed in the case
of $\tauR = 2.0\fm$.

These dips can be better understood by observing a heat map similar to
Fig.~\ref{Fig:corr3d.n-sigma-nu}.  Figure~\ref{Fig:correlation_n_3d} shows the
heat map of $\langle \tilde n(\Delta\etas) \tilde n(0)\rangle$ in the
$\etas$--$T$ plane for different baryon relation times $\tauR$.  The
qualitative behavior is similar to Fig.~\ref{Fig:corr3d.n-sigma-nu}, but the
structures are clearer.  In Fig.~\ref{Fig:correlation_n_3d}~(a), with the
vanishing baryon relaxation time $\tauR$, the negative pocket appearing on the critical
temperature $\Tc$ quickly diffuses, and the clear structure disappears before
the temperature reaches $150\MeV$.  However, with the finite relaxation times
in Fig.~\ref{Fig:correlation_n_3d}~(b) and~(c), the negative pocket travels to
larger $\Delta\etas$ regions at a finite speed keeping its shape.

The clearer structure in the $n$--$\nu$ system in
Fig.~\ref{Fig:correlation_n_3d} than in the $n$--$\sigma$--$\nu$ system in
Fig.~\ref{Fig:corr3d.n-sigma-nu} can be understood by the dispersion relations
of those systems.  Both the $n$--$\sigma$--$\nu$ and $n$--$\nu$ systems have
propagating modes at large $|\bm{k}|$ as observed in
Eqs.~\eqref{eq:dispersion.3x3-highk-prop}
and~\eqref{eq:dispersion.n-nu.general}, while only the $n$--$\sigma$--$\nu$
system has an additional diffusive mode~\eqref{eq:dispersion.3x3-highk-prop-2}
at large $|\bm{k}|$.  This means that the $n$--$\sigma$--$\nu$ system is more
diffusive than the $n$--$\nu$ system for small structures and explains that the
sharp structure is preserved more in the $n$--$\nu$ system.

\subsection{Temperature dependence of correlation function}
\label{sec:result.2pt.peak-vs-T}

The magnitude of the ultraviolet divergence of the equilibrium autocorrelation
functions at $\Delta\etas$ is related to the
susceptibilities~\eqref{eq:n-sigma.susceptibilities} (See,
e.g., Ref.~\cite{Asakawa:2015ybt}).  The autocorrelations in the present setup
of the expanding system do not reach the equilibrium ones, yet it is useful to
observe the magnitude of the peak at $\Delta\etas = 0$ and check the
differences between the dynamical systems and relaxation times.
We also
compare them to the equilibrium values.

\begin{figure}[htbp]
  \centering
  \includegraphics[width=0.48\textwidth,  bb=0 0 144 180]{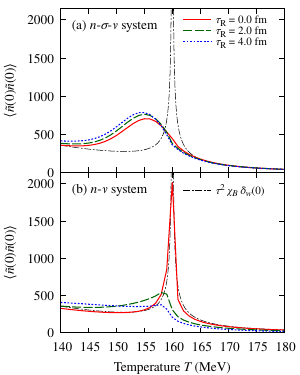}
  \caption{Magnitude of the peak at $\Delta\etas = 0$ in the autocorrelation of
  the baryon density fluctuation $\tilde n$, $\langle \tilde n (0) \tilde n (0)
  \rangle$, shown as a function of temperature $T$.  Panels~(a) and~(b) show
  the results for the $n$--$\sigma$--$\nu$ and $n$--$\nu$ systems,
  respectively.  The red solid line, the green dashed line, and the blue dashed
  line correspond to the vanishing relaxation time, $\tauR=2.0\fm$, and 4.0\fm,
  respectively.  }
  \label{Fig:correlation_T}
\end{figure}
\begin{figure}[htbp]
  \includegraphics[width=0.48\textwidth,  bb=0 0 144 93]{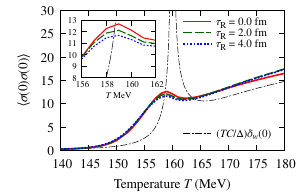}
  \caption{Magnitude of the peak at $\Delta\etas = 0$ in the autocorrelation
  $\langle \tilde \sigma(0) \tilde \sigma(0) \rangle$ in the
  $n$--$\sigma$--$\nu$ system, shown as a function of temperature $T$.  The
  line colors and styles are the same as Fig.~\ref{Fig:correlation_T}.  The
  inset shows the same quantity but with a finer resolution near the critical
  temperature $\Tc$.}
  \label{fig:sigma-correlation-vs-T}
\end{figure}

In Fig.~\ref{Fig:correlation_T}~(a) and~(b), we compare the autocorrelations
$\langle \tilde n(0) \tilde n(0) \rangle$ in expanding systems as functions of
the temperature $T$.  Panels~(a) and~(b) show the results in the
$n$--$\sigma$--$\nu$ and $n$--$\sigma$ systems, respectively.  They also
contain $n$--$\sigma$ and $\sigma$ systems, respectively, as the vanishing
baryon relation cases $\tauR = 0\fm$.  The red solid, green dashed, and blue
dotted lines show the results with different baryon relaxation times $\tauR$.  The
black thin dot-dashed line shows the equilibrium values of the peak magnitude,
\begin{align}
  \langle\tilde n(0)\tilde n(0)\rangle_\mathrm{eq}
    &= \tau^2 \langle n(0) n(0)\rangle_\mathrm{eq} \notag \\
    &= \tau^2 \chiB \delta_w(\Delta\etas = 0) = \frac{\tau \chiB}{S_\perp\sqrt{4\pi w^2}}\;,
\end{align}
where $\delta_w(\Delta\etas)$ is the smeared delta
function in Eq.~\eqref{eq:mode.n-sigma-nu-expanding.smeared-delta}.

We first notice significant qualitative differences between the
$n$--$\sigma$--$\nu$ and $n$--$\nu$ systems.  While peaks as functions of $T$
are formed near the critical temperature in the $n$--$\nu$ system in
Fig.~\ref{Fig:correlation_T}~(b), the peaks become round hills and shift toward
lower temperature in the $n$--$\sigma$--$\nu$ system in
Fig.~\ref{Fig:correlation_T}~(a).  In both systems, the positions of the peaks
are slightly shifted to a lower temperature also by the finite baryon
relaxation time.

In the $n$--$\sigma$--$\nu$ system shown in Fig.~\ref{Fig:correlation_T}~(a),
the hill becomes taller with a larger baryon relaxation time, which is because
the baryon density fluctuations do not immediately escape from the $\Delta\etas
\sim 0$ region due to the finite propagation speed ensured by the finite baryon
relaxation.  The same trends are observed in the $n$--$\nu$ system at the lower
temperature $T \lesssim 150\MeV$ in Fig.~\ref{Fig:correlation_T}~(b).
Conversely, near the critical temperature $\Tc$, a smaller baryon relaxation
time generates larger peak values in the $n$--$\nu$ system. This is because the
system responds to the singularity in the coefficients more quickly with a
smaller relaxation time.  The same trends are slightly seen in the
$n$--$\sigma$--$\nu$ just near $\Tc$.

In the $n$--$\nu$ system shown in Fig.~\ref{Fig:correlation_T}~(b), we see a
significant peak at $\Tc$ with the vanishing baryon relaxation time $\tauR =
0\fm$ as shown by the red solid line, which mostly reproduces the equilibrium
values for each temperature $T$ shown by the black dot-dashed line.  This means
that the dynamical effect is small in the present setup of the $n$ system.  The
introduction of the baryon diffusion $\nu$ makes the peak insignificant in the
$n$--$\nu$ system.  In particular, with $\tauR = 4.0\fm$ shown by the blue
dotted line in Fig.~\ref{Fig:correlation_T}~(b), the enhancement of the
fluctuations is mostly hidden in the background.  However, when we also
consider dynamical $\sigma$, the enhanced fluctuations remain finite with large
baryon relaxation times. Also, the enhancement remains at the lower temperature
$T\simeq 155\MeV$ than in the vanishing $\tauR$ case in
Fig.~\ref{Fig:correlation_T}~(b) with the divergent fluctuations.  This is
favorable for detecting the signal of critical fluctuations in experiments.

In the $n$--$\sigma$--$\nu$ system, we can also check the peak magnitude in the
autocorrelation of $\sigma$.  The equilibrium value is given by
\begin{align}
  \langle\sigma(0)\sigma(0)\rangle_\mathrm{eq}
    &= \frac{TC}{\Delta} \delta_w(0) = \frac{TC}\Delta \frac1{\tau S_\perp\sqrt{4\pi w^2}}\;.
\end{align}
In Fig.~\ref{fig:sigma-correlation-vs-T},
the dependence on the baryon relaxation time $\tauR$ is not large
because $\tauR$ affects the baryon sector and the effect on $\sigma$ is
indirect.  As shown in the inset, the $\sigma$-fluctuations become slightly
larger for a smaller $\tauR$ for the same reason as the baryon autocorrelations
on the critical temperature $\Tc$ in Fig.~\ref{Fig:correlation_T}.

\section{Conclusion}
\label{sec:conclusion}

We have constructed a dynamical model for the critical dynamics in an expanding
system.
During the dynamical evolution of the system in heavy-ion collisions,
the scale separation between non-critical and critical modes becomes nontrivial.
As an instructive attempt to quantify
the effects of mode coupling between non-critical and critical modes,
we coupled the baryon density fluctuation $n$ to the chiral condensate
fluctuation $\sigma$ and the diffusion current $\nu$ with the baryon relaxation
time $\tauR$.
Within this model, we analyzed (1+1)d spacetime evolution of
these fluctuations and their correlations.

We first showed the Green functions with the relaxation time $\tauR = 0.0$,
$2.0$, and $4.0\fm$.
We have confirmed that causality is satisfied with an appropriate value of
$\tauR = 4.0\fm$, which is consistent with the signal propagation speed based
on the dispersion relation.
Also, two peaks propagate to the $\pm\etas$ directions
with a finite
relaxation time. The finite baryon relaxation time is important for
causality and significantly affects the qualitative behavior of the system.
Next, we checked the behavior around the critical temperature, where each
propagating peak was found to split into two propagating peaks.
This is caused by
the divergence of kinetic coefficient $\lambda$ in our model,
which suggests
the importance of mode coupling and the need to go beyond the linear analysis.
With dynamical $\sigma$, the signal propagation speed becomes faster, but the
structures become less clear. These are understood from the dispersion
relation.
We then analyzed the correlation functions. We have found that the sharp
enhancement of the ultraviolet peak of the $n$-autocorrelation near the
critical temperature is smeared by dynamical $\sigma$, but it does not affect
the peak size in the later time.  We have also found that the effect in the
later stage becomes more significant for finite $\tauR$.  This means that the signal of
the critical temperature would be more preserved in the final state than the
prediction of the analyses without $\tauR$.

In the future, the model of critical dynamics should be implemented in a
state-of-the-art dynamical model of high-energy nuclear collisions, and the
qualitative and quantitative impact on the experimental observables should be
investigated in a more realistic setup. In finding experimental observables
sensitive to critical fluctuations,
the finite baryon relaxation time $\tauR$
should be taken into account
because it helps the signal be preserved in later
times.  The coupling to energy and momentum, and the nonlinear effects coming from
higher-order terms in the free energy functional~\cite{Nahrgang:2018afz} are
also important topics, which are left for future studies.

\section*{Acknowledgement}
The numerical calculations were performed on Supercomputer Yukawa-21 at Yukawa
Institute for Theoretical Physics (YITP) at Kyoto University.  This work was
supported by JSPS KAKENHI Grant Numbers JP18J22227 (A.S.), JP19K21881 (T.H. and
H.F.), and JP23K13102 (K.M.), and also supported by the National Natural
Science Foundation of China (NSFC) under Grant No.~11947236.

\appendix

\section{Susceptibilities \texorpdfstring{$\chiB(T)$ and $\chi^\mathrm{reg}(T)$}{χ\_B(T) and χ\textcircumflex reg(T)}, and diffusion constant \texorpdfstring{$D(T)$}{D(T)}}
\label{app:sakaida}

In this section, we summarize $\chiB(T)$, $\chi^\mathrm{reg}(T)$, and $D(T)$
originally given by Ref.~\cite{Sakaida:2017rtj} along with our choice of the
parameters.

The full susceptibility $\chiB(T)$ is defined to be a sum of the critical and
regular contributions, $\chi^\mathrm{cr}$ and $\chi^\mathrm{reg}$, as in
Eq.~\eqref{eq:param.A-in-chiB}.  For the critical part $\chi^\mathrm{cr}$, we
rely on the universality of the critical behavior and a mapping between the
Ising model variables $(r, H)$ and the QCD variables $(T,
\mu)$~\cite{Berdnikov:1999ph, Nonaka:2004pg}.  The variables $r$ and $H$ are
the reduced temperature and the magnetic field, respectively, in the Ising model.  In
this work, we assume a linear mapping between $(r, H)$ and $(T, \mu)$, and a
constant $r$ during the evolution.  The magnetic field at the Ising side, $H$,
is associated with the temperature $T$ by the relation,
\begin{align}
  \frac{T-T_{\rm c}}{\Delta T} = \frac{H}{\Delta H}\;.
\end{align}
The ratio $\Delta T/\Delta H$ controls the size of the critical region while
$r$ controls the distance of the evolution trajectory from the critical point.
The origin $(r,H)=(0,0)$ is the critical point, $\{(r,H) | r<0, H=0\}$ is the
first-order line, and $r>0$ is the crossover region.  In this study, we choose
$\Delta T/\Delta H = 10\ \text{MeV}$ and $r=10^{-4}$.

The critical part of the baryon number susceptibility is expressed as
\begin{align}
  \frac{\chi^\mathrm{cr}(r, H)}{\chi^\mathrm{H}}
  = c_\mathrm{c} \chi_M (r, H),
\end{align}
where $\chi^\mathrm{H}=1$ is the susceptibility in the hadronic phase, and
$c_\mathrm{c}=4$ is a constant.  We use a parametrization of the magnetic susceptibility
$\chi_M(r,H)$ of the Ising model:
\begin{align}
  \chi_M (r, H)
  &= \frac{m_0}{h_0}\frac{1}{R^{4/3}(3+2\theta^2)},
\end{align}
where $m_0= \sqrt{2/3}\, 2^{-\beta}$, $h_0 = m_0^\delta$, with the exponents
$(\beta,\delta)=(1/3, 5)$.  The coordinates $(R,\theta)$ are obtained from
$(r,H)$ by solving the following equations:
\begin{align}
  r &= R(1-\theta^2)\;, \\
  H &= h_0 R^{\beta\delta} \theta(3-2\theta^2)\;.
\end{align}

For the regular part of the susceptibility $\chi^\mathrm{reg}$,
we smoothly interpolate it between
the quark-phase susceptibility at high temperature $\chi^\mathrm{Q}_0$
and the hadronic one at low temperature $\chi^\mathrm{H}_0$:
\begin{align}
  \chi^\mathrm{reg} (T) &= \left[1-S(T)\right]\chi^\mathrm{H}_0 +
  S(T)\chi^\mathrm{Q}_0\;, \\ S(T) &= \frac{1}{2}\left[1 + \tanh
  \left(\frac{T-\Tc}{\delta T}\right)\right]\;,
\end{align}
where the width of the crossover is $\delta T = 10\ \text{MeV}$.
The values of $\chi^\mathrm{H}_0$ and $\chi^\mathrm{Q}_0$ are determined by
\begin{subequations}
\begin{align}
  \chi^\mathrm{H} &= \chi^\mathrm{cr}(T_\mathrm{f}) + \chi^\mathrm{H}_0\;,
  \label{eq:def.chiH0} \\
  \chi^\mathrm{Q} &= \chi^\mathrm{cr}(T_{0}) + \chi^\mathrm{Q}_0\;,
  \label{eq:def.chiQ0}
\end{align}
\end{subequations}
so that they reproduce $\chi^\mathrm{H}=1$ at the final temperature
$T_\mathrm{f} = 100\ \text{MeV}$ and $\chi^\mathrm{Q}=1/2$ at the initial
temperature $T_0 = 220\ \text{MeV}$.

The dynamic critical behavior of the QCD critical point shares the universality
class with the liquid-gas critical point~\cite{Son:2004iv, Fujii:2004jt,
Fujii:2004za}.  The kinetic coefficient $\lambda$ behaves as $\lambda\propto
\xi^{\chi_\lambda}$ in the critical region with the correlation length $\xi$
and the exponent $\chi_\lambda$.  In terms of the susceptibility, $\chi_M
\propto \xi^{2-\eta}$, we can write the singular part of the diffusion constant
as
\begin{align}
  D^{\rm cr}(T) &= d_c
    \left[\frac{\chi^\mathrm{cr}(T)}{\chi^\mathrm{H}}\right]^{-1 + \frac{\chi_\lambda}{2-\eta}}\;.
\end{align}
We use the results from a renormalization group calculation, $\eta = 0.04$ and
$\chi_\lambda = 0.916$~\cite{Hohenberg:1977ym}.  For the proportionality
constant $d_c$, we set $d_c=1\ \text{fm}$.

We express the total diffusion constant also as a sum of the critical and
regular parts:
\begin{align}
  \frac{1}{D(T)} = \frac{1}{D^\mathrm{cr}(T)} + \frac{1}{D^\mathrm{reg}(T)}\;.
\end{align}
The regular part of the diffusion coefficient is defined as
\begin{align}
  D^{\rm reg}(T)=[1-S(T)]D_0^\mathrm{H} + S(T) D_0^\mathrm{Q}\;,
\end{align}
where the parameter $D^\mathrm{Q}_0$ and $D^\mathrm{H}_0$ are determined by
\begin{subequations}
\begin{align}
  \frac1{D^\mathrm{H}} &= \frac1{D^\mathrm{cr}(T_\mathrm{f})} + \frac1{D^\mathrm{H}_0}\;, \\
  \frac1{D^\mathrm{Q}} &= \frac1{D^\mathrm{cr}(T_0)} + \frac1{D^\mathrm{Q}_0}\;,
\end{align}
\end{subequations}
to reproduce $D^\mathrm{H}=0.6$ at the final temperature $T_\mathrm{f}$ and
$D^\mathrm{Q}=2.0$ at the initial temperature $T_0$.

\section{Numerical method}
\label{app:numerical}
\newcommand{\etasi}[1]{\eta_{\mathrm{s},#1}}

We here describe the numerical method to solve the time evolution of the
$n$--$\sigma$--$\nu$ system in the $\tau$--$\etas$ coordinates and obtain the
time evolution of the $n$--$\sigma$ systems as the limit of $\tauR\to0$.  We
first write the equations of motion in the following form:
\begin{subequations}
\begin{align}
  \frac{d\tilde n}{d\tau} &= -\nabla\nu\;,
    \label{eq:numerical.eom-n} \\
  \frac{d\sigma}{d\tau} &= -\frac1{\tau_\sigma}(\sigma - Y_\sigma)\;,
    \label{eq:numerical.eom-sigma} \\
  \frac{d\nu}{d\tau} &= -\frac1{\tauR}(\nu - Y_\nu)\;,
    \label{eq:numerical.eom-nu}
\end{align}
\end{subequations}
where  $\nabla = \partial/\partial\etas$, $\tau_\sigma = 1/\Gamma A$, and
\begin{subequations}
\begin{align}
  Y_\sigma
    &= \tau_\sigma\biggl[- \Gamma B \frac{\tilde{n}}{\tau} + \xi_\sigma\biggr]\;, \\
  Y_\nu
    &= -\frac{1}{\tau} \biggl[(\tilde{\lambda}A + \lambda B) \nabla\sigma
      + (\tilde{\lambda}B + \lambda C) \frac{\nabla\tilde{n}}{\tau}\biggr] + \xi_\nu\;.
\end{align}
\end{subequations}

To solve the continuity equation~\eqref{eq:numerical.eom-n}, which is in the
divergence form, we employ the FVM (finite volume method)-type spatial
discretization: We split the $\etas$ coordinate into cells labeled by the index
$i$. Then, we define $\{\tilde n_i\}_i$ and $\{\sigma_i\}_i$ in the cells and
the diffusion current $\{\nu_{i+1/2}\}_i$ at the cell boundaries.  We consider
the equal widths of the cells, $\Delta\etas = 0.01$.  The quantities in the
cells, $\sigma_i$ and $\tilde n_i$, are specifically defined as the spatial
averages within the respective cells:
\begin{subequations}
\begin{align}
  \sigma_i(\tau) &:= \int_{\etasi{i-1/2}}^{\etasi{i+1/2}}\frac{d\etas}{\Delta\etas} \sigma(\tau, \etas)\;, \\
  \tilde n_i(\tau) &:= \int_{\etasi{i-1/2}}^{\etasi{i+1/2}}\frac{d\etas}{\Delta\etas} \tilde n(\tau, \etas)\;,
  \label{eq:numerical.def-n}
\end{align}
where $\etasi{i+1/2}$ denotes the position of the cell boundary between the $i$
and $(i+1)$th cells.  The diffusion current $\nu_{i+1/2}$ is simply defined as
the boundary value:
\begin{align}
  \nu_{i+1/2}(\tau) &:= \nu(\tau, \eta_{i+1/2})\;.
\end{align}
This set of the variables, $(\{\sigma_i\}_i, \{\tilde n_i\}_i,
\{\nu_{i+1/2}\}_i)$, specifies the state of the system.
\end{subequations}

\begin{subequations}
The integral form of the continuity equation~\eqref{eq:numerical.eom-n} for the
charge in the finite volume~\eqref{eq:numerical.def-n} is obtained by the
Stokes theorem:
\begin{align}
  &\tilde n_i(\tau + \Delta\tau) \notag \\
  &\quad = \tilde n_i(\tau) + \frac1{\Delta\etas}\int_\tau^{\tau+\Delta\tau} d\tau' [\nu_{i+1/2}(\tau') - \nu_{i-1/2}(\tau')] \notag \\
  &\quad \simeq \tilde n_i(\tau) + \frac1{\Delta\etas}\int_\tau^{\tau+\Delta\tau} d\tau' [\nu_{i+1/2}(\tau) - \nu_{i-1/2}(\tau)] \notag \\
  &\quad = n_i(\tau) + \frac{\Delta\tau}{\Delta\etas}[\nu_{i+1/2}(\tau) - \nu_{i-1/2}(\tau)]\;.
  \label{eq:numerical.update-n}
\end{align}
On the third line, we approximated $\nu_{i+1/2}(\tau') = \nu_{i+1/2}(\tau) +
\mathcal{O}(\tau-\tau')$ by $\nu_{i+1/2}(\tau)$. This results in the time
discretization error of $\mathcal{O}(\Delta\tau^2)$, so this is the first-order
time integrator.

To handle the stiffness of the relaxation
equations~\eqref{eq:numerical.eom-sigma} and~\eqref{eq:numerical.eom-nu} with
small values of $\tauR$ and $\tau_\sigma$, we exactly integrate the diagonal
parts for the finite time step $\Delta\tau$, i.e., the solution of the
relaxation equation is approximated as
\begin{align}
  &\nu_{i+1/2}(\tau + \Delta\tau) \notag \\
  &\quad = e^{-\frac{\Delta\tau}{\tauR}}\nu_{i+1/2}(\tau)
    + \frac1{\tauR}\int_\tau^{\tau + \Delta\tau}d\tau' e^{-\frac{\tau - \tau'}{\tauR}}Y_{\nu,i+1/2}(\tau') \notag \\
  &\quad \simeq e^{-\frac{\Delta\tau}{\tauR}}\nu_{i+1/2}(\tau)
    + \frac1{\tauR}\int_\tau^{\tau + \Delta\tau}d\tau' e^{-\frac{\tau - \tau'}{\tauR}}Y_{\nu,i+1/2}(\tau) \notag \\
  &\quad = e^{-\frac{\Delta\tau}{\tauR}}\nu_{i+1/2}(\tau)
    + (1 - e^{-\frac{\Delta\tau}{\tauR}}) Y_{\nu,i+1/2}(\tau)\;,
  \label{eq:numerical.update-nu}
\end{align}
where we replaced $Y_{\nu,i+1/2}(\tau')$ with $Y_{\nu,i+1/2}(\tau)$ on the
third line, which is an approximation consistent with
Eq.~\eqref{eq:numerical.update-n}.  We use the same update rule for the
relaxation of $\sigma_i$:
\begin{align}
  \sigma_i(\tau + \Delta\tau)
  &\simeq e^{-\frac{\Delta\tau}{\tau_\sigma}}\sigma_i(\tau) + (1 - e^{-\frac{\Delta\tau}{\tau_\sigma}}) Y_{\sigma,i}(\tau)\;.
  \label{eq:numerical.update-sigma}
\end{align}
\end{subequations}

In the limit of the small time step $\Delta\tau/\tauR$ and $\Delta\tau/\tauR$,
up to $\mathcal{O}(\Delta\tau)$,
Eqs.~\eqref{eq:numerical.update-nu} and~\eqref{eq:numerical.update-sigma} become
\begin{subequations}
\begin{align}
  & \nu_{i+1/2}(\tau + \Delta\tau) \notag \\
  & \quad = \nu_{i+1/2}(\tau) -\frac{\Delta\tau}{\tauR} [\nu_{i+1/2}(\tau) - Y_{\nu,i+1/2}(\tau)], \\
  & \sigma_i(\tau + \Delta\tau)
    = \sigma_i(\tau) -\frac{\Delta\tau}{\tau_\sigma} [\sigma_i(\tau) - Y_{\sigma,i}(\tau)]\;,
\end{align}
\end{subequations}
where the noise terms $\xi_{\nu,i+1/2}$ and $\xi_{\sigma,i}$ appear in
$Y_{\nu,i+1/2}(\tau)$ and $Y_{\sigma,i}(\tau)$.  This defines the It\^o type of
the stochastic differential equation.  It should be noted that the It\^o and
Stratonovich types give the same solution in the present setup because the
noise magnitude does not depend on the dynamical fields.

In the limit of $\tauR/\Delta\tau \to 0^+$, one observes that
Eq.~\eqref{eq:numerical.update-nu} reduces to $\nu_{i+1/2}(\tau + \Delta\tau) =
Y_{\nu,i+1/2}(\tau)$.  Therefore, the right-hand side of
Eq.~\eqref{eq:numerical.update-n} would be reduced to the second-order finite
difference as expected, but using the fields of the second previous step:
$\nu_{i+1/2}(\tau) = Y_{\nu,i+1/2}(\tau-\Delta\tau)$.  However, we expect it to
become the one using the previous step, which is the natural update rule for
the $n$--$\sigma$ system.  We modify the update rule by introducing the
operator splitting of the time evolution: $\{\nu_{i+1/2}\}_i$ are updated using
Eq.~\eqref{eq:numerical.update-nu} first, and then $\{\sigma_i\}_i$ and
$\{\tilde n_i\}_i$ are updated by Eqs.~\eqref{eq:numerical.update-n}
and~\eqref{eq:numerical.update-sigma} using the new values of
$\{\nu_{i+1/2}\}_i$.  This ensures that the scheme is reduced to that of the
diffusion equation written by the finite difference of the current field in the
limit of $\tauR/\Delta\tau \to 0^+$.

The terms $Y_{\nu,i+1/2}$ and $Y_{\sigma,i}$ in
Eqs.~\eqref{eq:numerical.update-nu} and~\eqref{eq:numerical.update-sigma} are
evaluated using the fields at the corresponding positions.  The spatial
derivatives in $Y_{\nu,i+1/2}$ are evaluated by the finite difference:
\begin{subequations}
\begin{align}
  (\nabla\tilde n)_{i+1/2}(\tau) &= \frac{\tilde n_{i+1}(\tau) - \tilde n_{i}(\tau)}{\Delta\etas}\;, \\
  (\nabla\sigma)_{i+1/2}(\tau) &= \frac{\sigma_{i+1}(\tau) - \sigma_{i}(\tau)}{\Delta\etas}\;.
\end{align}
\end{subequations}
The noise fields appearing in $Y_{\nu,i+1/2}$ and $Y_{\sigma,i}$ are smeared by
the Gaussian width $w$:
\begin{subequations}
\begin{align}
  \xi_{\sigma,i} &= \sum_{k=-N}^{N} \frac1{\sqrt{2\pi w^2}}e^{-\frac{(k\Delta\eta)^2}{2w^2}}\xi_{\sigma,i+k}^0\;, \\
  \xi_{\nu,i+1/2} &= \sum_{k=-N}^{N} \frac1{\sqrt{2\pi w^2}}e^{-\frac{(k\Delta\eta)^2}{2w^2}}\xi_{\nu,i+k+1/2}^0\;,
\end{align}
\end{subequations}
where $N = 10$ is used as a sufficiently large number, and the bare noise
fields are generated by
\begin{subequations}
\begin{align}
  \xi_{\sigma,i}^0 &= \sqrt{\frac{2T\Gamma}{\tau\Delta\etas S_\perp\Delta\tau}} \hat{w}_i\;, \\
  \xi_{\nu,i+1/2}^0 &= \sqrt{\frac{2T\lambda}{\tau\Delta\etas S_\perp\Delta\tau}} \hat{w}_{i+1/2}\;,
\end{align}
\end{subequations}
with $\hat{w}_i$ and $\hat{w}_{i+1/2}$ being independent pseudorandom numbers
following the standard normal distribution.  The correlation between
$\xi_\sigma$ and $\xi_\nu$ is not considered because we assume vanishing
$\tilde\lambda$ in the present analysis.

\section{Green's function in response to $\nu$-impulse}
\label{app:green.nu-impulse}

\begin{figure*}[htbp]
  \centering
  \includegraphics[width=0.99\textwidth]{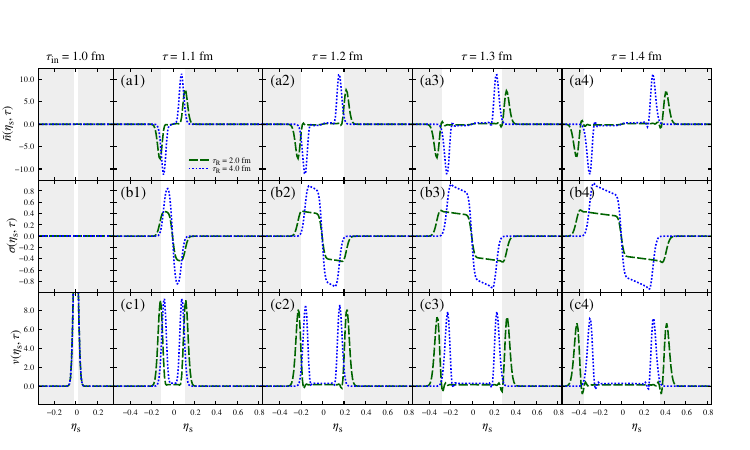}
  \includegraphics[width=0.99\textwidth]{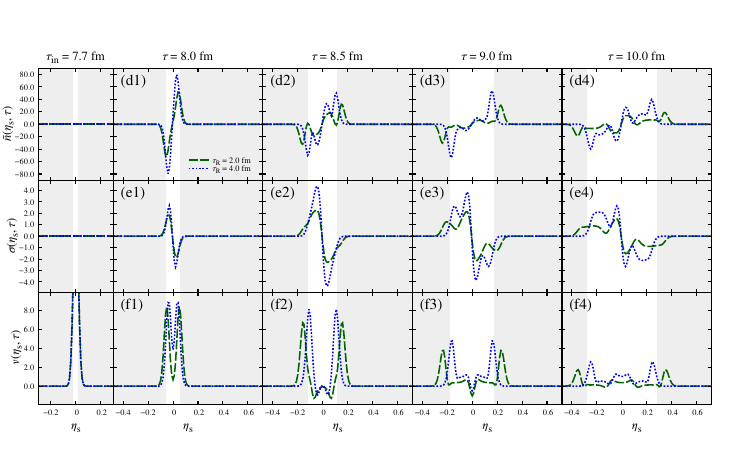}
  \caption{Spacetime evolution of $\tilde n(\etas, \tau)$, $\sigma(\etas,
  \tau)$, and $\nu(\etas, \tau)$ in response to the initial impulse in $\nu$ in
  the $n$--$\sigma$--$\nu$ system.  $\tauR$, $\tauI$, and line types are the
  same as in Fig.~\ref{Fig:evolution_baryon}.
  The case for the vanishing baryon relaxation time $\tauR=0\fm$ is not plotted
  because $\nu$ is not a dynamical degree of freedom in this case and cannot be
  independently specified in the initial condition.}
  \label{Fig:evolution_nu}
\end{figure*}

\begin{figure*}[htbp]
  \centering
  \includegraphics[width=0.99\textwidth]{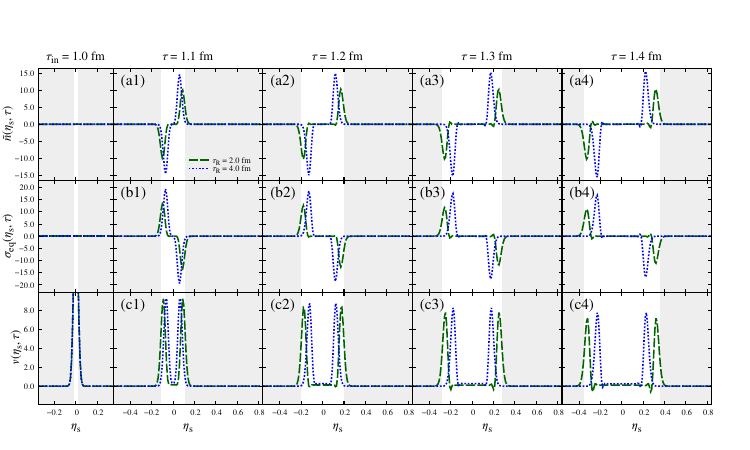}
  \includegraphics[width=0.99\textwidth]{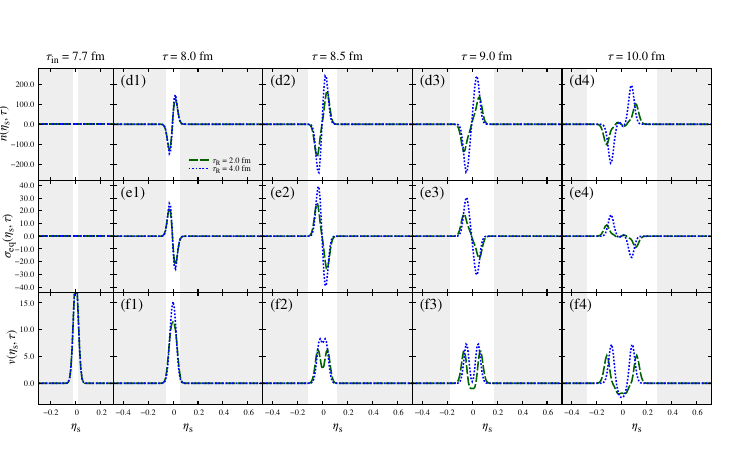}
  \caption{Spacetime evolution of $\tilde n(\etas, \tau)$, $\sigma(\etas,
  \tau)$, and $\nu(\etas, \tau)$ in response to the initial impulse in $\nu$ in
  the $n$--$\nu$ system.  $\tauR$, $\tauI$, and line types are the same as in
  Fig.~\ref{Fig:evolution_baryon}.}
  \label{Fig:evolution.n-nu.nu}
\end{figure*}

In Sec.~\ref{sec:Green}, we checked the Green's functions in response to the
impulses in $\tilde n$ and $\sigma$.  However, when the baryon relaxation time
$\tauR$ is finite, the full set of the Green's functions includes also the
response to the impulse in the baryon diffusion current $\nu$ because $\nu$ is
also dynamical.  In this section, we supplement the results for the response to
the $\nu$-impulse for both the $n$--$\sigma$--$\nu$ and $n$--$\nu$ systems.

The initial condition for the $\nu$-impulse is given by
\begin{align}
  \nu(\etas, \tauI) &= \frac1{\sqrt{2\pi w^2}} e^{-\etas^2/2w^2}\;,
  \label{eq:green.impulse.nu}
\end{align}
with $w = 0.02$.  The other fields are initialized to vanish: $\tilde n(\etas,
\tauI) = \sigma(\etas, \tauI) = 0$.  The results are shown only for $\tauR =
2.0$ and $4.0\fm$ because $\nu$ is not dynamical in the vanishing $\tauR$ case.
In the same manner as
Figs.~\ref{Fig:evolution_baryon}--\ref{Fig:evolution.n-nu.n},
Figs.~\ref{Fig:evolution_nu} and~\ref{Fig:evolution.n-nu.nu} show the time
evolutions in the $n$--$\sigma$--$\nu$ and $n$--$\nu$ systems, respectively.

In Fig.~\ref{Fig:evolution_nu}~(a1), we see that $\tilde n$ has two
antisymmetric peaks around the origin.  Since the local diffusion current
pushes the baryon charge density in the positive $\etas$ direction, the baryon
density locally increases in a positive $\etas$, while it decreases in a
negative $\etas$.  Then, $\tilde n$ induces the opposite sign of $\sigma$ as
seen in Fig.~\ref{Fig:evolution_nu}~(b1) to create two peaks of $\sigma$.  The
$\nu$ peak also splits into two peaks as in Fig.~\ref{Fig:evolution_nu}~(c1).
As time passes, those structures grow into the $\pm\etas$ directions with the
finite speed defined by $v$~\eqref{eq:dispersion.3x3-highk-prop}.  The peaks of
$\tilde n$ and $\nu$ shift their positions to larger $\etas$, while the peaks
of $\sigma$ expand into larger $\etas$ to form plateaux.

In the lower three rows (d)--(f) of Figs.~\ref{Fig:evolution_nu}, like in the
cases of other impulses in Figs.~\ref{Fig:evolution_baryon}
and~\ref{Fig:evolution_sigma}, we see that new structures are created when the
system passes the critical temperature $\Tc$. As a result, the structures
become less clear than in the evolutions far from the critical point in
Figs.~\ref{Fig:evolution_nu}~(a)--(c).

Figure~\ref{Fig:evolution.n-nu.nu} shows the responses to the $\nu$-impulse for
the $n$--$\nu$ system.  We observe the same characteristics as in
Fig.~\ref{Fig:evolution_baryon}: The chiral condensate fluctuation $\sigma$ is
proportional to $-n$ since it is determined by $\sigma_\mathrm{eq} = -
(B/A)(\tilde n/\tau)$.  We also see that the propagation speed becomes slower
compared to Fig.~\eqref{Fig:evolution_nu}.

\bibliography{References}
\end{document}